\documentclass[aps,pre,preprint,showpacs]{revtex4-2}
\usepackage{color}
\usepackage{colordvi}
\usepackage{epsfig}
\usepackage{bm,amssymb}
\usepackage{mathtools}

\begin{document}

\title{
The distribution of the number of cycles \\
in directed and undirected random 2-regular graphs 
}

\author{Ido Tishby}

\affiliation{Racah Institute of Physics, 
The Hebrew University, Jerusalem, 9190401, Israel}

\author{Ofer Biham}

\affiliation{Racah Institute of Physics, 
The Hebrew University, Jerusalem, 9190401, Israel}

\author{Eytan Katzav}

\affiliation{Racah Institute of Physics, 
The Hebrew University, Jerusalem, 9190401, Israel}

\author{Reimer K\"uhn}

\affiliation{Mathematics Department, King's College London, 
Strand, London WC2R 2LS, UK}

\begin{abstract}

We present analytical results for the distribution of the number of cycles
in directed and undirected random 2-regular graphs (2-RRGs) 
consisting of $N$ nodes.
In directed 2-RRGs each node has one inbound link and one outbound link,
while in undirected 2-RRGs each node has two undirected links.
Since all the nodes are of degree $k=2$, the resulting networks
consist of cycles.
These cycles exhibit a broad spectrum of lengths,
where the average length of the shortest cycle in a random network instance scales
with $\ln N$, 
while the length of the longest cycle scales with $N$.
The number of cycles varies between different network instances
in the ensemble,
where the mean number of cycles $\langle S \rangle$ 
scales with $\ln N$.
Here we present exact analytical results for the distribution 
$P_N(S=s)$
of the number of cycles $s$
in ensembles of directed and undirected 2-RRGs,
expressed in terms of the Stirling numbers of the first kind.
In both cases 
the distributions converge to a Poisson distribution in the large $N$ limit.
The moments and cumulants of $P_N(S=s)$ are also calculated.
The statistical properties of directed 2-RRGs are equivalent to 
the combinatorics of cycles in random permutations of $N$ objects.
In this context our results recover and extend known results.
In contrast, the statistical properties of cycles in undirected 2-RRGs 
have not been studied before.

\end{abstract}


\maketitle

\newpage

\section{Introduction}

Random networks (or graphs) consist of a set of $N$ nodes that are connected
to each other by edges in a
way that is determined by some random process.
They provide a useful conceptual framework for the study of a large variety
of systems and processes in science, technology and society
\cite{Bollobas2001,Dorogovtsev2003,Havlin2010,Estrada2011,Barrat2012,Hofstad2016,Latora2017,Newman2018,Dorogovtsev2022}. 
The structure of a random network can be characterized by the degree distribution
$P(k)$.
Here we focus on a special class of random networks,
called random regular graphs (RRGs), which exhibit
degenerate degree distributions of the form

\begin{equation}
P(k) = \delta_{k,c},
\end{equation}

\noindent
where $\delta_{k,k'}$
is the Kronecker delta and
$c \ge 1$ is an integer.
While the degrees of all the nodes in these networks are the same,
their connectivity is random and uncorrelated.
In that sense, the RRG is a special case of the class of configuration model
networks, which exhibit a specified degree distribution $P(k)$, but no
degree-degree correlations
\cite{Bollobas1980,Molloy1995,Molloy1998,Newman2001}.
While RRGs with $c=1$ consist of isolated dimers, RRGs with $c \ge 3$
form a giant component.
Thus, RRGs with $c=2$ are a marginal case, separating between the subcritical
regime of $c < 2$ and supercritical regime of $c > 2$.
Note that unlike some other configuration model networks that exhibit a coexistence
of a giant component and finite tree components above the percolation transition,
in RRGs with $c \ge 3$ the giant component  
encompasses the whole network.

RRGs with $c=2$, referred to as 
random 2-regular graphs (2-RRGs),
consist of isolated cycles of various lengths. 
The lengths of these cycles are not determined
by the topology, but by entropic considerations.
An important distinction is between directed 2-RRGs, in which each node
has one inbound link and one outbound link, and undirected
2-RRGs, in which each node has two undirected links.
In both cases the cycles can be
considered as isolated components of the network, where the length of each cycle is
equal to the size of the network component that consists of this cycle.

In this paper we present exact analytical results for the distribution of
the number of cycles in ensembles of directed and undirected 2-RRGs  
that consist of $N$ nodes. 
The results are expressed in terms of the Stirling numbers of the first kind.
We first calculate the joint probability distribution of  
cycle lengths.
This is done by mapping the directed and
undirected 2-RRGs into combinatorically equivalent permutation problems.
From the joint distribution of cycle lengths we extract the distribution 
$P_N(S=s)$ of the number of
cycles in random instances of directed and undirected 2-RRGs.
In both cases $P_N(S=s)$ converges to a Poisson distribution in the large $N$ limit.
The moments and cumulants of $P_N(S=s)$ are also calculated.
The similarities and differences between the results obtained for the directed and undirected
2-RRGs are discussed.
The statistical properties of directed 2-RRGs are equivalent to 
the combinatorics of cycles in random permutations of $N$ objects.
In this context our results recover and extend known results.
In contrast, the statistical properties of cycles in undirected 2-RRGs 
have not been studied before.
The results presented in this paper are derived specifically for $c=2$
and do not apply to the more general case of RRGs with $c \ge 3$.

The paper is organized as follows. 
In Sec. II we present the directed and undirected 2-RRGs.
The joint distributions of cycle lengths are presented in Sec. III.
In Sec. IV we calculate the distribution $P_N(S=s)$ of the number of cycles.
In Sec. V we calculate the moments and cumulants of $P_N(S=s)$.
The results are discussed in Sec. VI and summarized in Sec. VII.

\section{Random 2-regular graphs}

In both directed and undirected 2-RRGs, each node has two
links and the resulting network consists of a set of cycles. 
In the directed case each node has one inbound link and
one outbound link, while in the undirected case  each node has 
two undirected links. Below we briefly present the properties of
directed and undirected 2-RRGs and their construction.

\subsection{Directed 2-RRGs}

To construct a directed 2-RRG one first assembles $N$ nodes such that each node
has one inbound stub and one outbound stub.
During the network construction process, at each time 
step one picks a random outbound stub and a 
random inbound stub among the remaining open stubs
and connects them to each other. 
This process is repeated $N$ times, until no open stubs remain. 
This procedure follows the standard construction process of directed configuration model networks,
in which pairs of outbound and inbound
stubs rather than pairs of nodes are selected for connection.
The resulting ensemble of networks obtained from this procedure is referred to as
stub-labeled graphs
\cite{Fosdick2018}.
During the construction process, directed
chains of nodes of different lengths are formed. 
In case that the open outbound stub of one chain is selected to connect to the open inbound stub of
another chain, they are connected and form a longer chain.
In case that the outbound and the inbound stubs at both ends of the same chain are
selected, their connection closes the chain and turns it into a cycle.
Once a chain of nodes becomes a cycle it does not connect to other
nodes and its structure remains unchanged.
At the end of the construction, the whole network
consists of cycles of different lengths.

\begin{figure}
\begin{center}
\includegraphics[width=6cm]{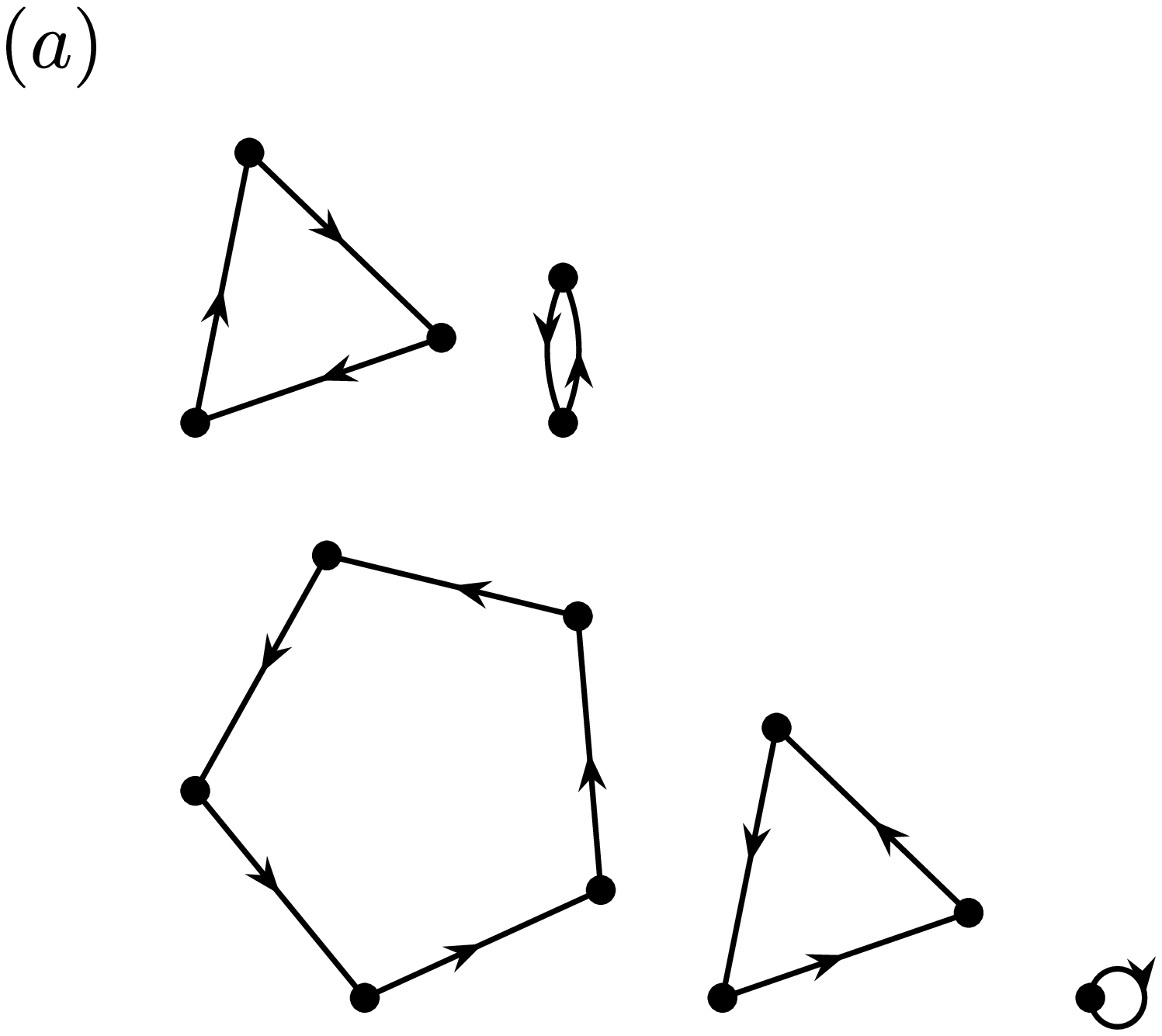} 
\hspace{0.5in}
\includegraphics[width=6cm]{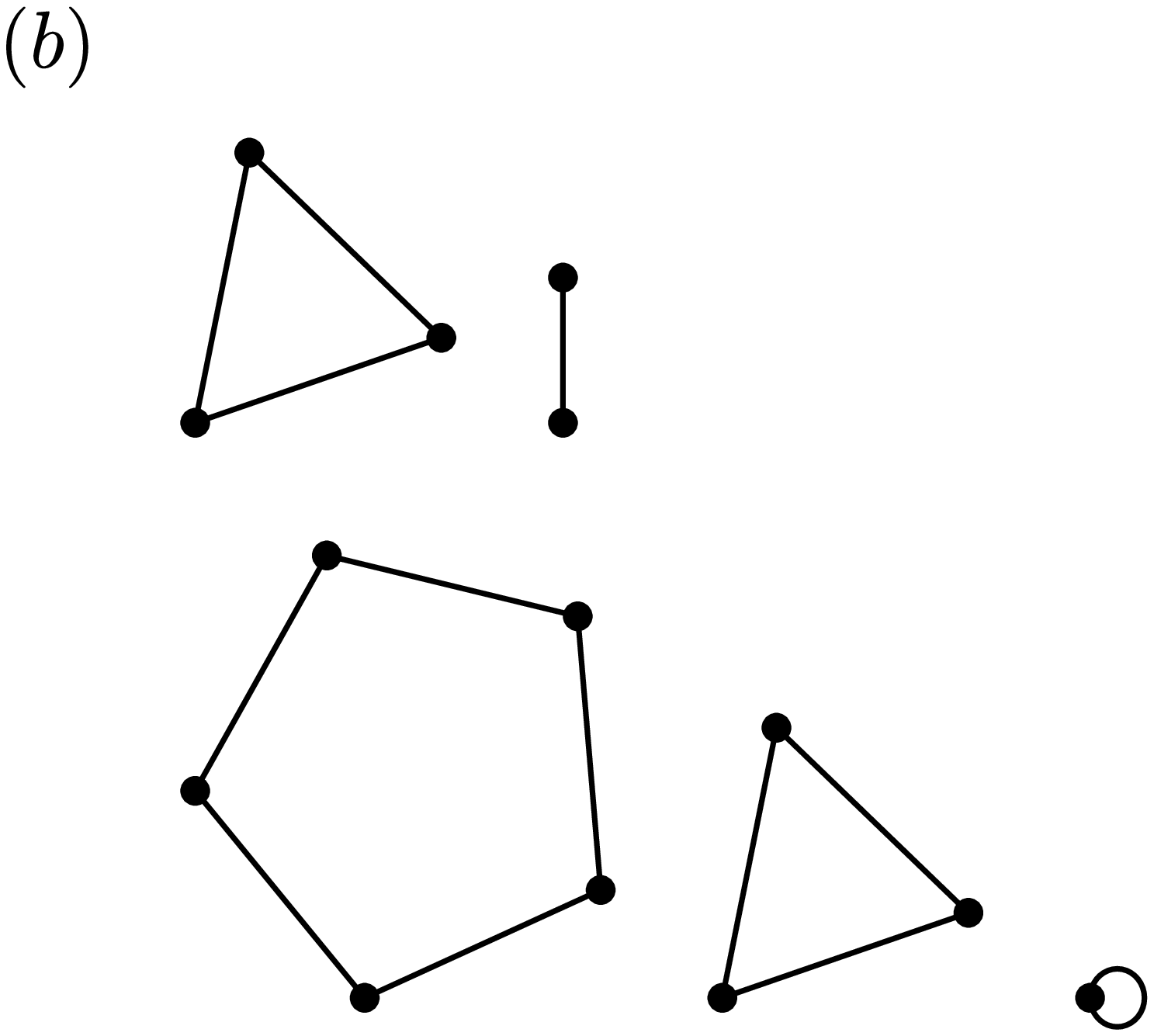} 
\end{center}
\caption{
Illustration of a single instance of a 2-RRG of $N=14$ nodes,
which consists of cycles of lengths $s=1, 2, 3, 3$ and $5$,
in the directed case (a) and in the undirected case (b).
}
\label{fig:1}
\end{figure}

In Fig. \ref{fig:1}(a)  we present 
a single instance of a directed 2-RRG of $N=14$ nodes,
which consists of cycles of lengths $s=1, 2, 3, 3$ and $5$.
In the 2-RRGs considered here
we allow the outbound and inbound stubs
of the same node to be connected to each other. In such case, one
obtains a self-loop of length $\ell=1$. We also allow the connection of a
pair of outbound and inbound stubs, which belong to nodes that are
already connected.
In such case, the resulting cycle is of length $\ell=2$.
This choice simplifies the analysis.

Consider a directed 2-RRG consisting of $N$ distinguishable nodes, which are marked
by the labels $i=1,2,\dots,N$. 
In the construction of such a network the first selected inbound stub 
has $N$ possible outbound stubs to which it may connect.
The second selected inbound stub has only $N-1$ outbound stubs
to which it may connect. Continuing this process we conclude that
there are $N!$ possible ways to connect the $N$ nodes.
In fact, the combinatorics of directed 2-RRGs 
consisting of $N$ nodes is equivalent to the permutation
problem of $N$ objects  
\cite{Shepp1966,Arratia1992,Flajolet1990}.
This permutation problem can be 
described by a line of $N$ cells labeled by 
$i=1,2,\dots,N$, and a corresponding set of $N$ labeled balls.
The number of ways to distribute the balls between the cells, one 
ball in each cell, is

\begin{equation}
R_{\rm D} = N!,
\label{eq:R_D}
\end{equation}

\noindent
where each arrangement of the balls in the cells corresponds
to a specific network instance. In this representation, a ball $i'$ located
in cell $i$ represents a directed link from $i$ to $i'$. Similarly, a ball $i''$
located in cell $i'$ represents a directed link from $i'$ to $i''$. One can
follow these directed links until the cell in which ball $i$ resides is reached
and the cycle is closed. 
Repeating this process for all the balls and cells provides the structure of
cycles associated with the specific permutation.
The length of each cycle is given by the number of balls included
in the cycle.

The statistical properties of cycles in random permutations 
of $N$ objects
have been
studied extensively
\cite{Golomb1964,Bona2012,Golomb2017}.
In particular, the properties of the longest cycle in each permutation
received much attention. 
It was found that the expectation value of the length of the longest
cycle in a random permutation of $N$ objects
is equal to $\lambda N$, where $\lambda = 0.6243...$ is the
Golomb-Dickman constant
\cite{Dickman1930,Golomb1998,Finch2003}.
Interestingly, this constant plays a role in the prime factorization 
of a random integer. 
More specifically, it was found that the asymptotic average number
of digits in the largest prime factor of a random $N$-digit number
is $\lambda N$
\cite{Dickman1930,Knuth1976,Finch2003}.
In the other extreme, the average length of the shortest cycle
in a random permutation of $N$ objects
was also calculated and was found to be
$e^{-\gamma} \ln N$, where $\gamma=0.5772...$ is the Euler-Mascheroni
constant
\cite{Shepp1966,Finch2003}.

\subsection{Undirected 2-RRGs}

To construct an undirected 2-RRG  one 
first assembles $N$ nodes such that each node is connected to two
undirected stubs.
At each time step one selects a random pair of stubs among the remaining open stubs and connects them
to each other. 
This process is repeated $N$ times, until no open stubs remain. 
This procedure follows the standard construction process of undirected configuration model networks,
in which pairs of stubs rather than pairs of nodes are selected for connection
\cite{Bollobas1980,Molloy1995,Molloy1998,Newman2001}.
The resulting ensemble of networks obtained from this procedure is referred to as
stub-labeled graphs
\cite{Fosdick2018}.
Initially, one obtains linear chains of nodes of increasing 
lengths that eventually close and form cycles.
Here we consider undirected 2-RRGs in which
we allow the two stubs
of the same node to be connected to each other and form a self-loop of length $\ell=1$. 
We also allow the connection of a
pair of  stubs which belong to nodes that are
already connected, resulting in a cycle of length $\ell=2$.
In Fig. \ref{fig:1}(b) we present 
a single instance of an undirected 2-RRG of $N=14$ nodes,
which consist of cycles of lengths $s=1, 2, 3, 3$ and $5$.

Consider an ensemble of undirected 2-RRGs consisting of $N$ nodes.
In the construction of such network the first selected stub has
$2N-1$ other stubs to which it may connect. The second selected stub
has $2N-3$ other stubs to which it may connect. Continuing this process
we conclude that there are  

\begin{equation}
R_{\rm U} = (2N-1)!!.
\label{eq:Ncu}
\end{equation}

\noindent
possible ways to construct such network,
where $m!!$ is the double factorial of $m$.
The statistical properties of the resulting ensemble of networks can be mapped to the 
combinatorial problem described below.
Consider a set of $2N$ balls, such that for each value of $i=1,2,\dots,N$
there are two identical balls on which the label $i$ is marked. 
The two balls labeled by a given value of $i$ represent the two stubs of node $i$.
In addition, there are $N$ unlabeled boxes such that in 
each box there is room for two balls.
The $2N$ balls are then distributed uniformly at random in the $N$ boxes, where
each box contains two balls.
Each pair of balls that resides in the same box represents one edge
of the network. For example, if a ball labeled by $i$ and a ball labeled by $i'$
are in the same box, it means that there is an edge between the nodes
$i$ and $i'$. Similarly, if the other ball labeled by $i'$ is in the same box
with a ball labeled $i''$, it means that there is an edge between the nodes
$i'$ and $i''$. One can follow this chain until reaching the box in which
the second ball labeled by $i$ resides, thus closing the cycle.
The number of possible ways to distribute the $2N$ balls in the $N$ cells
is given by

\begin{equation}
R_{\rm U} = \frac{ (2N)! }{ 2^{N} N!},
\label{eq:Ncu0}
\end{equation}

\noindent
where the numerator accounts for the number of permutations
of $2N$ balls, the $N!$ term in the denominator accounts for the
number of permutations of the $N$ identical cells, and the term
$2^{N}$ accounts for the fact 
that the order in which the two balls are placed in
each cell is unimportant. 
Note that $(2N)!=(2N-1)!!(2N)!!$ 
and $2^N N!=(2N)!!$.
These two identities establish the equivalence between
Eqs. (\ref{eq:Ncu}) and (\ref{eq:Ncu0}).

\section{The joint distribution of cycle lengths}

Both versions of the 2-RRG, with directed and undirected links,
consist of closed cycles of different lengths. 
The configuration of cycles in a given network instance 
can be described by the sequence of cycle lengths,
$\ell_1,\ell_2,\dots,\ell_s$, 
where 
$1 \le \ell_i \le N$,
$s$ is the number of cycles in the given network instance
and

\begin{equation}
\sum_{i=1}^s \ell_i = N. 
\label{eq:sumL}
\end{equation}

\noindent
For convenience and uniqueness, we order the lengths
in increasing order, such that 
$\ell_1 \le \ell_2 \le \dots \le \ell_s$.

Another way to describe the configuration of cycles in a given 
network instance is in the form
$\{ g_{\ell} \}_{\ell=1}^N =  \{ g_1,g_2,\dots,g_N \}$, 
where $g_{\ell}$ is the number of cycles of length $\ell$.
The $g_{\ell}$'s satisfy the condition

\begin{equation}
\sum_{\ell=1}^N \ell g_{\ell} =N,
\label{eq:sumN}                         
\end{equation}

\noindent
which is equivalent to Eq. (\ref{eq:sumL}).
The number of cycles in such a network instance can be expressed by

\begin{equation}
s = \sum_{\ell=1}^N g_{\ell}.                          
\end{equation}

\noindent
The joint distribution of cycle lengths in an ensemble of 2-RRGs consisting of $N$ nodes
is denoted by $P_N(\{ G_{\ell} \} = \{ g_{\ell} \})$, under the condition of 
Eq. (\ref{eq:sumN}). For convenience we use the notation $P_N(\{ g_{\ell} \})$.
Below we consider the joint distributions of
the cycle lengths in directed and undirected 2-RRGs.

\subsection{Joint distribution of cycle lengths in directed 2-RRGs}

Consider an ensemble of directed 2-RRGs that consist of $N$ nodes.
The number of configurations of the form
$\{ g_{\ell} \}$
is given by

\begin{equation}
N(\{ g_{\ell} \}) = N!  \prod_{\ell=1}^N  \frac{1}{   {\ell}^{g_{\ell}} g_{\ell}! }
\delta_{\sum \ell g_\ell , N}.
\end{equation}

\noindent
To understand the first term in the denominator, consider a cycle of length $\ell$. 
Such cycle exhibits $\ell$ cyclic permutations.
This yields the first term in the denominator,
which is raised to the power $g_{\ell}$ to account for
the number of cycles of length $\ell$. 
The term $g_{\ell}!$ accounts for the permutations
between the $g_{\ell}$ degenerate cycles of length $\ell$, which correspond to the same
configuration, while the Kronecker delta imposes the condition of Eq. (\ref{eq:sumN}).
Dividing the number of configurations 
$N(\{ g_{\ell} \})$ by the total number
of configurations $R_D$, 
given by Eq. (\ref{eq:R_D}),
we obtain the
joint probability distribution of cycle lengths.
It is given by

\begin{equation}
P_N(\{ g_{\ell} \}) =  \prod_{\ell=1}^N  \frac{1}{   {\ell}^{g_{\ell}} g_{\ell}! } 
\delta_{\sum \ell g_\ell , N}.
\label{eq:Pju}
\end{equation}

\noindent
It can be shown that this probability distribution is normalized, namely

\begin{equation}
\sum_{g_1} \sum_{g_2} \dots \sum_{g_N}
P_N(\{ g_{\ell} \}) = 1,
\end{equation}

\noindent
where the summation is over all the configurations of 
$\{ g_{\ell} \}$ that satisfy Eq. (\ref{eq:sumN}).

\subsection{Joint distribution of cycle lengths in undirected 2-RRGs}

Consider an ensemble of undirected 2-RRGs of $N$ nodes.
The number of configurations of the form
$\{ g_{\ell} \}$ is

\begin{equation}
N(\{ g_{\ell} \}) =  N! \prod_{\ell=1}^N 
\frac{2^{(\ell-1)g_{\ell}}}{\ell^{g_{\ell}} g_{\ell}!}
\delta_{\sum \ell g_\ell , N}.
\label{eq:N(g)}
\end{equation}

\noindent
This result can be understood in terms of the analogous combinatorial problem
described above, which consists of $N$ pairs of identical balls and $N$ unlabled boxes.
Inserting two random balls in each box, the
factor of $N!$ accounts for the number of
permutations of the $N$ pairs of indices marked on the balls.
To account for the other factors, consider a cycle of length $\ell$.
There are $2^{\ell}$ possible ways to exchange the $\ell$ pairs of identical balls.
However, due to the cyclic structure there is also a factor of $1/2$, because in each
cycle the lables marked on the balls can be listed either in the clockwise direction or
in the counterclockwise direction.
Taking this into account yields the factor
of $2^{ \ell-1 }$ in the numerator.
This factor is raised to the power $g_{\ell}$ to account for the
fact that there are $g_{\ell}$ cycles of length $\ell$.
In the denominator, the $\ell^{g_{\ell}}$ term accounts for
the number of cyclic permutations of the indices in all the cycles
of length $\ell$, while the $g_{\ell}!$ term accounts for the
number of permutations of $g_{\ell}$ degenerate cycles
of the same length.
The probability that a random network instance will
have a cycle structure given by $\{ g_{\ell} \}$
is given by

\begin{equation}
P_N(\{ g_{\ell} \}) = \frac{ N(\{ g_{\ell} \}) }{ R_U }.
\label{eq:Pglu0}
\end{equation}

\noindent
Inserting $N(\{ g_{\ell} \})$ from Eq. (\ref{eq:N(g)}), and $R_{\rm U}$
from Eq. (\ref{eq:Ncu})
into Eq. (\ref{eq:Pglu0}), 
we obtain

\begin{equation}
P_N(\{ g_{\ell} \}) = 
\frac{ N! }{ (2N-1)!! } \prod_{\ell=1}^N 
\frac{2^{(\ell-1)g_{\ell}}}{\ell^{g_{\ell}} g_{\ell}!}
\delta_{\sum \ell g_\ell , N}.
\label{eq:Pglu}
\end{equation}

\noindent
Using the relation
$2^N=2^{\sum_{\ell=1}^{N}\ell g_{\ell}}$,
we obtain

\begin{equation}
P_N(\{ g_{\ell} \}) = 
\frac{ (2N)!! }{ (2N-1)!! } \prod_{\ell=1}^N 
\frac{1}{(2\ell)^{g_{\ell}} g_{\ell}!}
\delta_{\sum \ell g_\ell , N}.
\label{eq:Pjd}
\end{equation}

\noindent
This expression differs from the corresponding result for directed 2-RRGs
in two ways: it has a factor of $(2\ell)^{g_{\ell}}$ in the denominator instead of $\ell^{g_{\ell}}$
and there is a pre-factor that is required in order to maintain the normalization.
The factor of $2^{g_{\ell}}$ in the denominator means that
as the number of cycles is increased the configuration becomes exponentially
less probable than the corresponding configuration of the directed 2-RRG.

\section{The distribution of the number of cycles}

The probability $P_N(S=s)$ that a random network instance
includes $s$ cycles can be calculated by summing up over all the
combinations of $\{ g_{\ell} \}$ that consist of $s$ cycles, namely

\begin{equation}
P_N(S=s) = \sum_{g_1,\dots,g_N} P_N( \{ g_{\ell} \}) \delta_{ \sum_{\ell} g_{\ell},s },
\label{eq:PNSs0}
\end{equation}

\noindent
where the configurations $\{ g_{\ell} \}$ satisfy the condition of Eq. (\ref{eq:sumN}).
Below we calculate the  distribution $P_N(S=s)$ for the directed and undirected 2-RRGs.

\subsection{Distribution of the number of cycles in directed 2-RRGs}

Inserting the expression for $P_N( \{ g_{\ell} \} )$ from Eq. (\ref{eq:Pju})
into Eq. (\ref{eq:PNSs0}), we obtain

\begin{equation}
P_N(S=s) = \sum_{g_1,\dots,g_N \ge 0} 
\prod_{\ell=1}^N
\frac{1}{   \ell^{g_{\ell}} g_{\ell} !  }  
\delta_{\sum \ell g_\ell , N}
\delta_{ \sum_{\ell} g_{\ell},s }.
\label{eq:PNSsD1}
\end{equation}

\noindent
For the analysis below it is convenient to perform a change of variables
from $g_{\ell}$, $\ell=1,2,\dots,N$ to $\ell_i$, $i=1,2,\dots,s$.
This transformation is based on the identity

\begin{equation}
\sum_{g_1,\dots,g_N \ge 0} 
\frac{1}{g_1! g_2! \dots g_N!} \delta_{ \sum_{\ell} g_{\ell},s }
=\frac{N^s}{s!} =
\frac{1}{s!} \sum_{\ell_1,\ell_2,\dots,\ell_s = 1}^{N} 1,
\end{equation}

\noindent
which is a result of the multinomial theorem
(equation 26.4.9 in Ref. \cite{Olver2010}).
Inserting the constraints that 
$\sum_{\ell=1}^N \ell g_{\ell} = N = \sum_{i=1}^s \ell_i$,
obtained from Eqs. (\ref{eq:sumL}) and (\ref{eq:sumN})
we obtain the identity

\begin{equation}
\sum_{g_1,\dots,g_N \ge 0} 
\frac{1}{g_1! g_2! \dots g_N!} \delta_{ \sum_{\ell} g_{\ell},s }
\delta_{\sum \ell g_\ell , N}
=
\frac{1}{s!} \sum_{\ell_1,\ell_2,\dots,\ell_s = 1}^{N} 
\delta_{ \sum_i \ell_i,N }
\label{eq:covgell}
\end{equation}

\noindent
Using this transformation, we express Eq. (\ref{eq:PNSsD1})
in the form

\begin{equation}
P_N(S=s) = 
\frac{1}{s!}
\sum_{\ell_1,\dots,\ell_{s}=1}^N
\frac{1}{ \ell_1 \ell_2 \dots \ell_{s} }
\delta_{ \sum_i \ell_i,N }.
\label{eq:Pnc}
\end{equation}

\noindent
Below we use the discrete Laplace transform,
which is related to the one-sided Z-transform and to the starred transform
\cite{Phillips2015},
to evaluate the right hand side of
Eq. (\ref{eq:Pnc}).
We denote the sum on the right hand side of Eq. (\ref{eq:Pnc}) by

\begin{equation}
f_s(N) = \sum_{   \ell_1,\dots,\ell_s \ge 1   }
\frac{1}{ \ell_1 \ell_2 \dots \ell_s } \delta_{\sum_{i=1}^s \ell_i,N}.
\label{eq:f(N)}
\end{equation}

\noindent
The discrete Laplace transform of $f_s(N)$ is given by

\begin{equation}
\widehat f_s(z) = \sum_{N=0}^{\infty} z^{N} f_s(N).
\label{eq:Lf(N)}
\end{equation}

\noindent
Inserting $f_s(N)$ from Eq. (\ref{eq:f(N)})
into Eq. (\ref{eq:Lf(N)}), we obtain

\begin{equation}
\widehat f_s(z) = \sum_{   \ell_1,\dots,\ell_s \ge 1   }
\frac{1}{ \ell_1 \ell_2 \dots \ell_s } 
\sum_{N=0}^{\infty} z^{N}
\delta_{ \sum_{i=1}^s \ell_i,N},
\label{eq:f(N)2}
\end{equation}

\noindent
where

\begin{equation}
\sum_{N=0}^{\infty} z^{N}
\delta_{\sum_{i=1}^s \ell_i,N} = z^{ \sum_{i=1}^s \ell_i }.
\label{eq:f(N)3}
\end{equation}

\noindent
Decomposing the multiple summation 
in Eq. (\ref{eq:f(N)2})
into a product of sums over the
$\ell_i$'s, we obtain

\begin{equation}
\widehat f_s(z) = \left(  \sum_{\ell=1}^{\infty} \frac{z^{\ell}}{\ell}  \right)^s.
\label{eq:hatf}
\end{equation}

\noindent
Carrying out the summation, we obtain

\begin{equation}
\widehat f_s(z) = \left[ - \ln \left(  1-z \right) \right]^s.
\label{eq:hatf}
\end{equation}

\noindent
The next step is to apply the inverse discrete Laplace transform on $\widehat f_s(z)$ 
to obtain $f_s(N)$. To this end we use identity
26.8.8 from Ref. \cite{Olver2010}, which is given by

\begin{equation}
\frac{ [\ln (1+y)]^k }{k!} = \sum_{n=0}^{\infty} s(n,k) \frac{y^n}{n!},
\label{eq:2688}
\end{equation}

\noindent
where $s(n,k)$ is the Stirling number of the first kind.
These Stirling numbers can be expressed in the form

\begin{equation}
s(n,k) =
(-1)^{n-k}
\bigg[
\begin{array}{l}
n     \\
k    
\end{array} 
\bigg], 
\label{eq:s(n,k)}
\end{equation}

\noindent
where
$\bigg[
\begin{array}{l}
n     \\
k    
\end{array} 
\bigg]$
is the unsigned Stirling number of the first kind
\cite{Olver2010}.
Inserting $s(n,k)$ from Eq. (\ref{eq:s(n,k)}) into Eq. (\ref{eq:2688}), we obtain

\begin{equation}
[\ln (1+y)]^k   = k! \sum_{n=0}^{\infty} (-1)^{n-k}
\bigg[
\begin{array}{l}
n     \\
k    
\end{array} 
\bigg]  \frac{y^n}{n!}.
\label{eq:2688b}
\end{equation}

\noindent
Inserting $y=-z$ into Eq. (\ref{eq:2688b})
we rewrite $\widehat f_s(z)$ in the form

\begin{equation}
\widehat f_s(z) = s! \sum_{n=0}^{\infty} \frac{1}{n!} 
\bigg[
\begin{array}{l}
n     \\
s    
\end{array} 
\bigg]
z^{n}.
\label{eq:fsx1}
\end{equation}

\noindent
To obtain the inverse discrete Laplace transform of $\widehat f_s(z)$ we use the fact
that

\begin{equation}
{\cal L}^{-1} \left\{ z^{ n } \right\} (N) = \delta_{N,n}.
\label{eq:Lm1}
\end{equation}

\noindent
Applying the inverse discrete Laplace transform on Eq. (\ref{eq:fsx1}), 
we obtain

\begin{equation}
f_s(N) = \frac{s!}{N!} 
\bigg[
\begin{array}{l}
N     \\
s    
\end{array} 
\bigg].
\label{eq:fN}
\end{equation}

\noindent
Inserting $f_s(N)$ from Eq. (\ref{eq:fN}) into Eq. (\ref{eq:Pnc}),
we obtain

\begin{equation}
P_N(S=s) = \frac{1}{N!} 
\bigg[
\begin{array}{l}
N     \\
s    
\end{array} 
\bigg].
\label{eq:PNSs}
\end{equation}

\noindent
The normalization of the distribution $P_N(S=s)$ can be confirmed using
identity 26.8.29 in Ref. \cite{Olver2010}.
In the context of permutations, the result expressed by Eq. (\ref{eq:PNSs})
implies that 
$\bigg[
\begin{array}{l}
N     \\
s    
\end{array} 
\bigg]$
counts the number of permutations with precisely $s$ cycles
among the $N!$ permutations of $N$ objects.
This is consistent with the combinatorial interpretation
of the unsigned Stirling number of the first kind
\cite{Olver2010}.
The cumulative distribution of the number of cycles is given by

\begin{equation}
P_N(S \le s) = \frac{1}{N!} 
\sum_{s'=1}^s
\bigg[
\begin{array}{l}
N     \\
s'    
\end{array} 
\bigg].
\label{eq:PNSsCum}
\end{equation}

\begin{figure}
\begin{center}
\includegraphics[width=7cm]{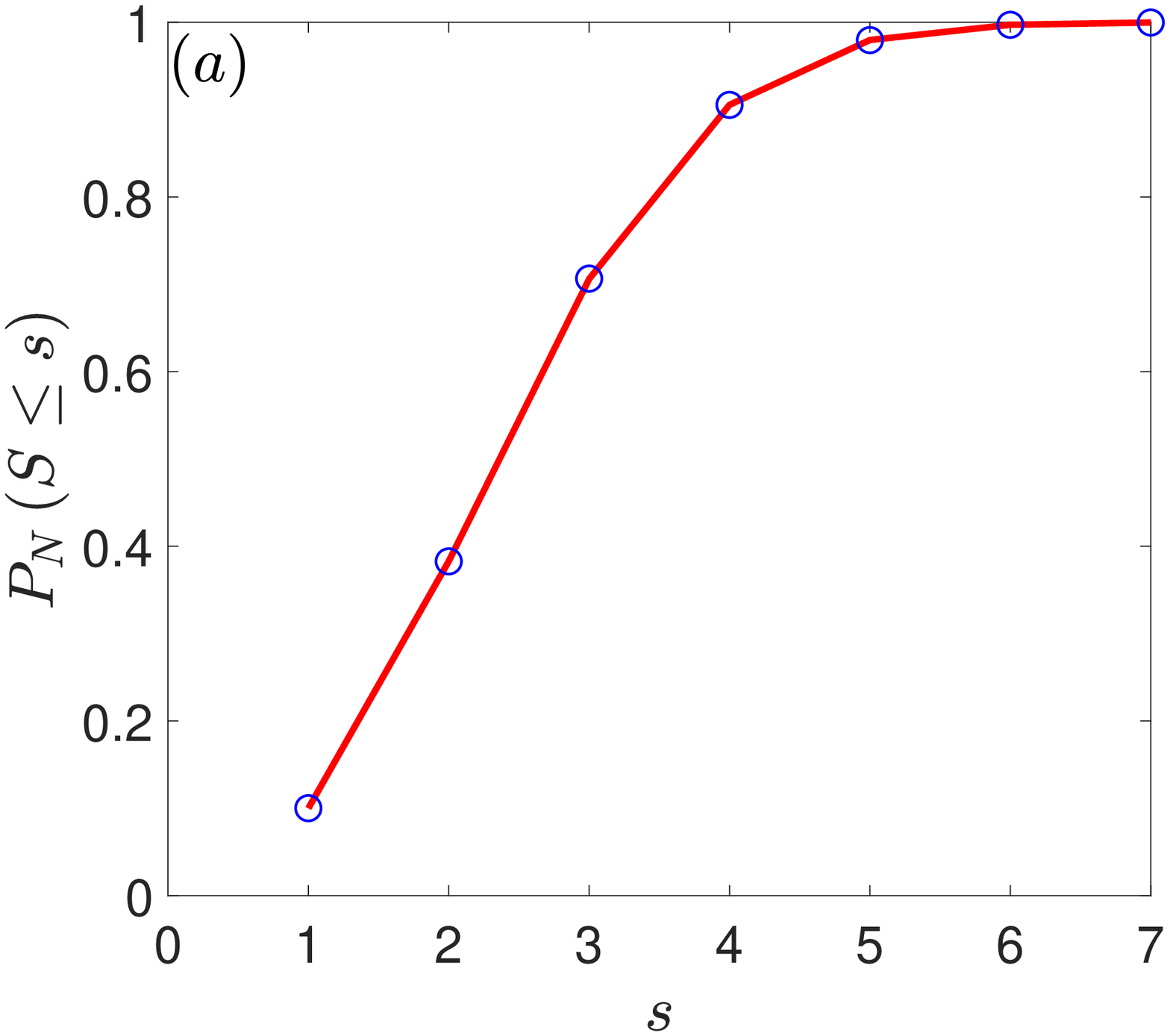} 
\hspace{0.5in}
\includegraphics[width=7cm]{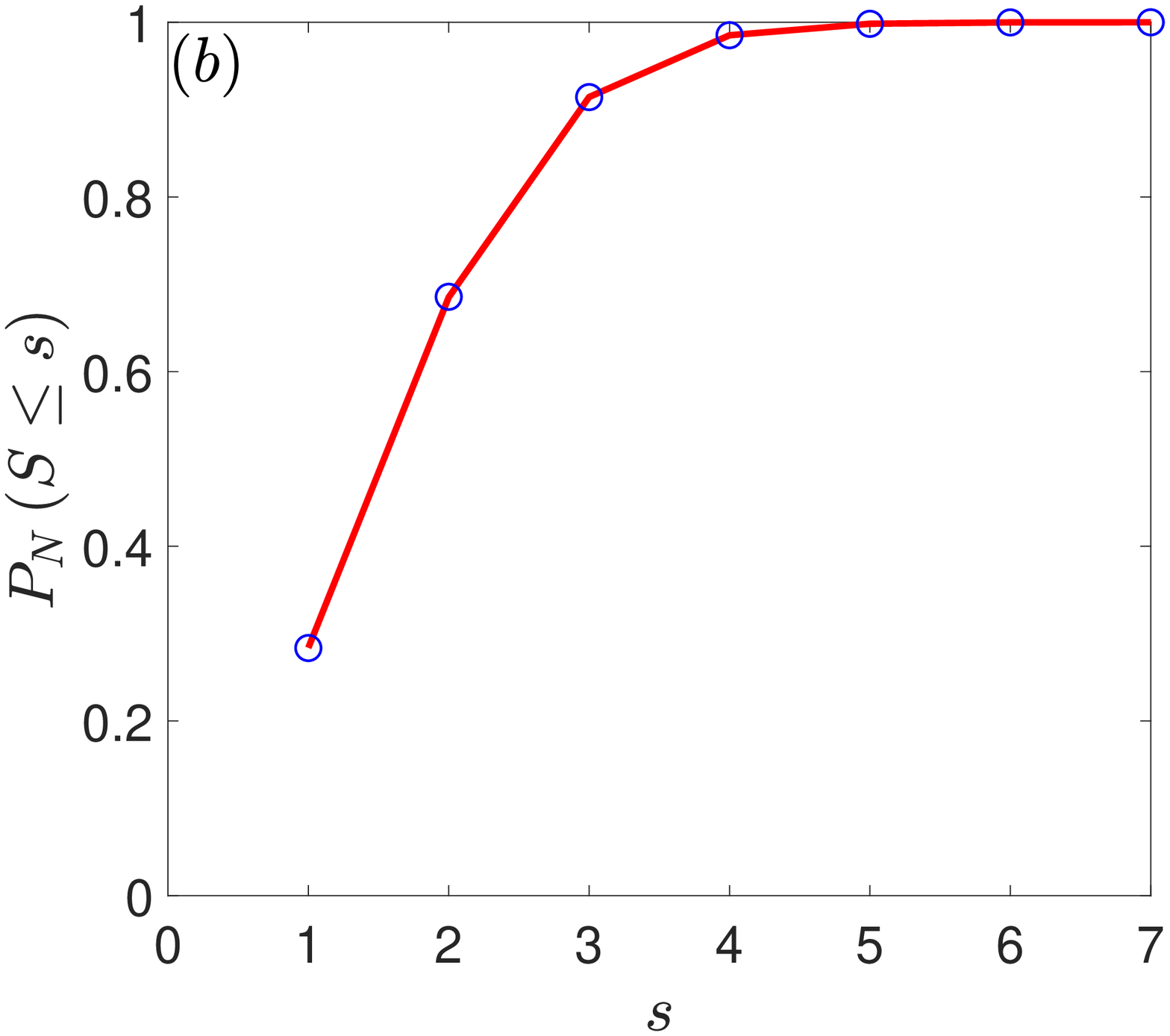} 
\end{center}
\caption{
Exact analytical results (solid lines) for the cumulative
distribution of the number of cycles in a directed 2-RRG (a)
and an undirected 2-RRG (b) of $N=10$ nodes,
obtained from Eqs. (\ref{eq:PNSsCum}) and (\ref{eq:PNSsUD_Cum}), 
respectively. The analytical results are in very good agreement with
the results obtained from computer simulations (circles).
}
\label{fig:2}
\end{figure}

In Fig. \ref{fig:2}(a) we present exact
analytical results (solid line) for the cumulative
distribution $P_N(S \le s)$ of the number of cycles in a directed 2-RRG 
that consists of $N=10$ nodes,
obtained from Eq. (\ref{eq:PNSsCum}). 
The analytical results are found to be in very good agreement with the
results obtained from computer simulations (circles).

While Eq. (\ref{eq:PNSs}) provides an exact result for $P_N(S=s)$, it is useful to 
express this distribution in terms of more elementary functions.
This is possible in the asymptotic limit of large $N$.
The limit of $N \gg 1$ corresponds to the limit of $z \rightarrow 1^{-}$
of the discrete Laplace transform $\widehat f_s(z)$ 
\cite{Schlomilch1852,Titchmarsh1939}.
If one replaces the variable $z$ by $e^{-x}$, then this limit corresponds to $x \ll 1$
in $\widehat f_s(e^{-x})$.
In this limit the expression for $\widehat f_s \left( e^{-x} \right)$ in Eq. (\ref{eq:hatf}) can be approximated by

\begin{equation}
\widehat f_s \left( e^{-x} \right) \simeq \left( - \ln x  \right)^s.
\label{eq:f_sx}
\end{equation}

\noindent
In order to calculate the inverse Laplace transform of $\widehat f_s \left( e^{-x} \right)$,
we use the relation

\begin{equation}
{\cal L}^{-1} \left[ \frac{1}{x^{\nu}} (- \ln x)^s \right]
= \left( \frac{d}{d \nu} \right)^s \left[ \frac{ N^{\nu-1} }{ \Gamma(\nu) } \right],
\label{eq:invLaplace}
\end{equation}

\noindent
where $\Gamma(\nu)$ is the Gamma function
\cite{Olver2010}.
The inverse Laplace transform of Eq. (\ref{eq:f_sx}) is obtained
by taking the limit $\nu \rightarrow 0$ in Eq. (\ref{eq:invLaplace}).
To this end we use the
general Leibnitz rule
(equation 1.4.12 in Ref. \cite{Olver2010})

\begin{equation}
\left( \frac{d}{d \nu} \right)^s \left[ \frac{ N^{\nu - 1}}{\Gamma(\nu)} \right]
=
\sum_{i=0}^s \binom{s}{i}
\left[ \left( \frac{d}{d \nu} \right)^{s-i} N^{\nu-1} \right]
\left[ \left( \frac{d}{d \nu} \right)^{i} \left( \frac{1}{\Gamma(\nu)} \right) \right].
\label{eq:binom}
\end{equation}

\noindent
Below we evaluate the derivatives that appear in the two terms 
on the right hand side of Eq. (\ref{eq:binom}).
The derivative in the first term is given by

\begin{equation}
\left( \frac{d}{d \nu} \right)^i N^{\nu - 1} = N^{\nu-1} (\ln N)^i.
\end{equation}

\noindent
Thus, for $\nu \rightarrow 0^{+}$ we obtain

\begin{equation}
\lim_{\nu \rightarrow 0^{+}}
\left[
\left( \frac{d}{d \nu} \right)^i N^{\nu - 1} 
\right]
= \frac{ (\ln N)^i }{N}.
\end{equation}

\noindent
The derivative in the second term on the right hand side of Eq. (\ref{eq:binom})
is denoted by

\begin{equation}
h_i = \lim_{\nu \rightarrow 0^{+}} 
\left\{
\left( \frac{d}{d \nu} \right)^i 
\left[ \frac{1}{\Gamma(\nu)} \right]
\right\}.
\end{equation}

\noindent
By its definition, $h_i$ is the coefficient of the $i$'th power
of $\nu$ in the Taylor expansion of $1/\Gamma(\nu)$ around
$\nu = 0$, namely

\begin{equation}
\frac{1}{\Gamma(\nu)} =
\sum_{i=1}^{\infty}
\frac{h_i}{i!} \nu^i.
\end{equation}

\noindent
This expansion is often written as a power-series of the form
\cite{Wrench1968,Wrench1973}

\begin{equation}
\frac{1}{\Gamma(\nu)} =
\sum_{i=1}^{\infty}
a_i \nu^i,
\label{eq:invGamma}
\end{equation}

\noindent
where $a_i = h_i/i!$.
The first two coefficients are given by
$a_1=1$ and $a_2 = \gamma$,
where $\gamma$ is the Euler-Mascheroni constant
\cite{Olver2010}.
Higher order coefficients can be obtained from the recursion equation

\begin{equation}
a_i = \frac{1}{i-1} \left[ a_2 a_{i-1} - \sum_{j=2}^{i-1} (-1)^j \zeta(j) a_{i-j} \right],
\label{eq:ai1}
\end{equation}

\noindent
where $\zeta(j)$ is the Riemann zeta function.
The coefficients $a_i$ for
$i=3,4,\dots,10$, 
obtained from Eq. (\ref{eq:ai1}),
are presented in Table I.
These coefficients can also be obtained from the integral representation
\cite{Ahmed2014}

\begin{table}
\caption{The coefficients $a_i$, $i=1,2,\dots,10$, 
which appear in the power series of the reciprocal gamma function
given by Eq. (\ref{eq:ai1})}
\begin{tabular}{|   r   |     r    |}
\hline \hline
\ $i$ \ \  & $a_i$  \ \ \ \ \ \  \ \    \\
\hline
\  1 \ \  & 1   \ \ \ \ \ \ \ \ \ \ \ \ \ \ \    \             \\
\  2 \ \ & \ \    0.5772156649  \ \  \   \\
\  3 \ \ & \ \  -  0.6558780715 \ \  \   \\
\  4 \ \ & \ \ -  0.0420026350  \ \  \   \\
\  5 \ \ & \ \    0.1665386114   \ \  \    \\
\  6 \ \ & \ \  -  0.0421977345  \ \ \    \\
\  7 \ \ & \ \  -  0.0096219715  \ \  \   \\
\  8 \ \ & \ \    0.0072189432    \ \  \   \\
\  9 \ \ & \ \  -  0.0011651675   \ \  \  \\
\  10 \ \ & \ \ - 0.0002152416  \ \  \   \\
 \hline
\end{tabular}
\end{table}

\begin{equation}
a_n = \frac{(-1)^n}{\pi n!} \int_0^{\pi}
e^{-t} {\rm Im} \left[ (\ln t - i\pi)^n \right] dt.
\label{eq:ai2}
\end{equation}

\noindent
Using this notation we obtain

\begin{equation}
f_s(N) \simeq \frac{s!}{N} \sum_{i=1}^{s}
a_i \frac{ (\ln N)^{s - i} }{(s-i)!},
\label{eq:f(N)4}
\end{equation}

\noindent
which leads to

\begin{equation}
P_N(S=s) \simeq \frac{1}{N} \sum_{i=1}^{s}
a_i \frac{ (\ln N)^{s - i} }{(s-i)!}.
\label{eq:Psd}
\end{equation}

\noindent
Note that for sufficiently large $N$ the sum is dominated by the first 
few terms for two reasons. First, apart from the first few terms,
the coefficients $a_i$ become negligibly small.
Second, the power of $\ln N$ decreases as $i$ is increased.

The cumulative distribution of the number of cycles is given by

\begin{equation}
P_N(S \le s) = 
\frac{1}{N}
\sum_{s'=1}^{s} 
\sum_{i=1}^{s'}
a_i \frac{ (\ln N)^{s' - i} }{(s' - i)!}.
\label{eq:PsdCum}
\end{equation}

\begin{figure}
\begin{center}
\includegraphics[width=7cm]{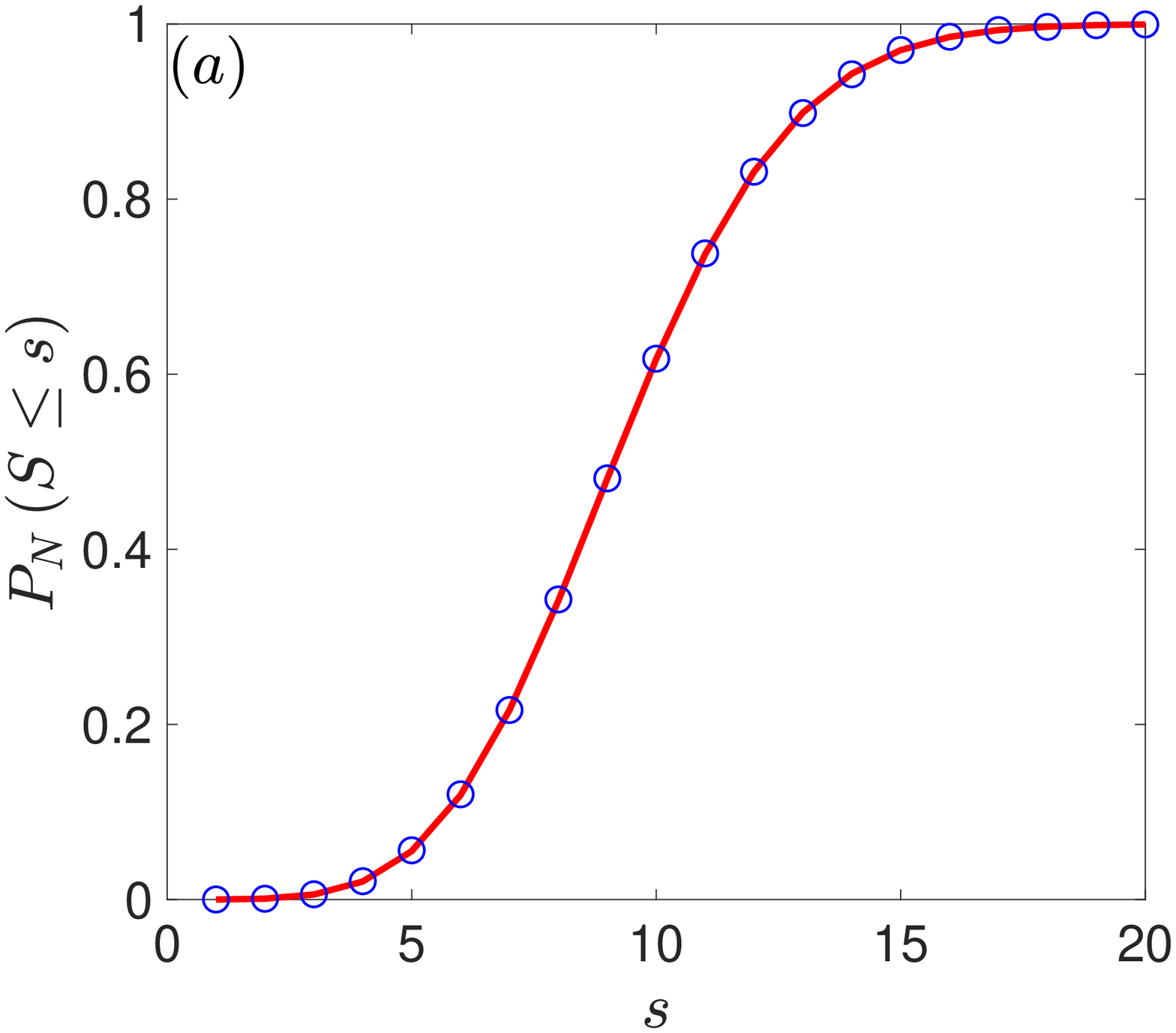} 
\hspace{0.5in}
\includegraphics[width=7cm]{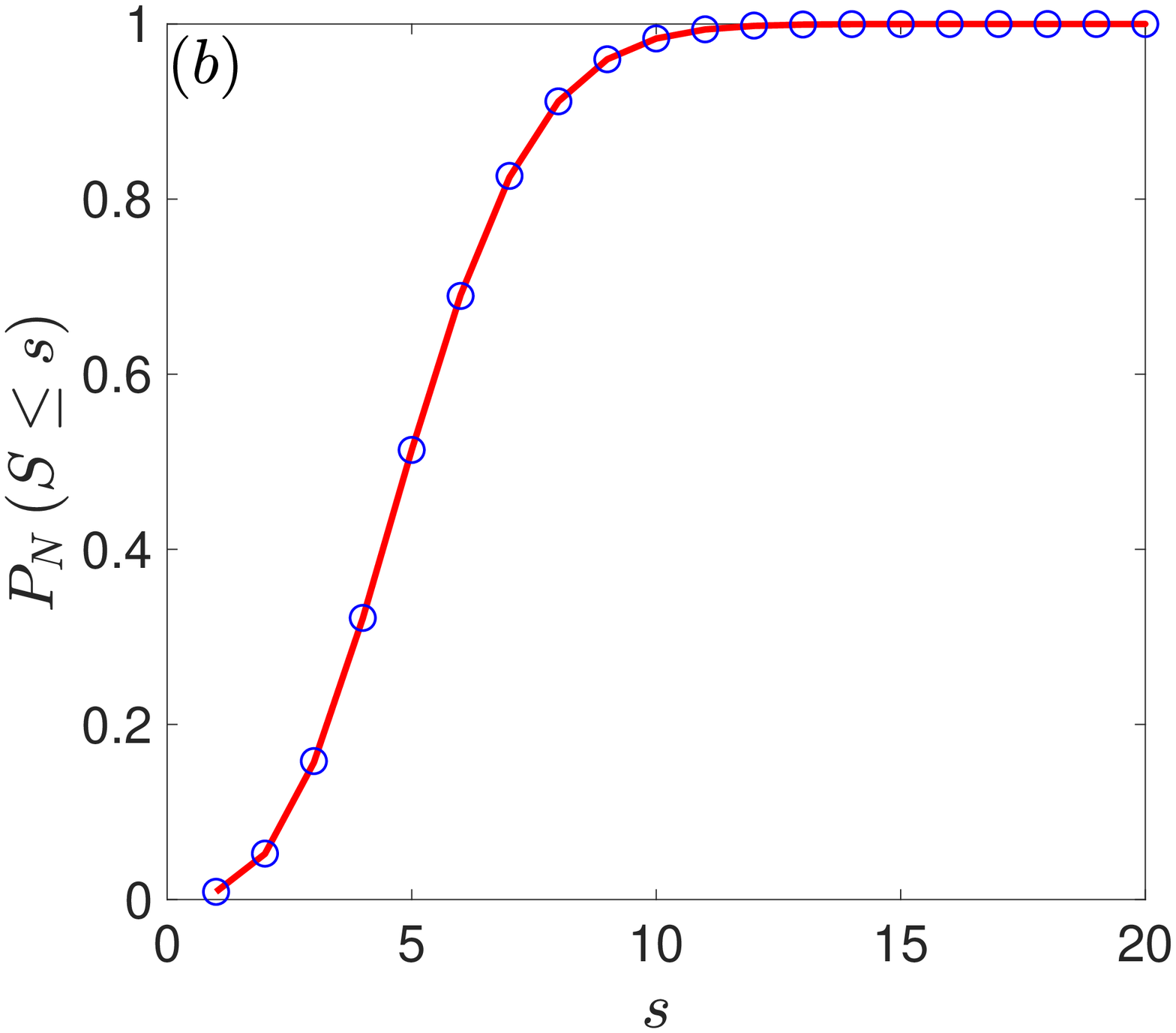} 
\end{center}
\caption{
Analytical results (solid lines) for the cumulative
distribution of the number of cycles in a directed 2-RRG (a)
and an undirected 2-RRG (b) of $N=10^4$ nodes,
obtained from Eqs. (\ref{eq:PsdCum}) and (\ref{eq:PsuCum}), 
respectively. The analytical results are in very good agreement with
the results obtained from computer simulations (circles).
}
\label{fig:3}
\end{figure}

\noindent
In Fig. \ref{fig:3}(a) we present
analytical results (solid line) for 
the large $N$ approximation of
the cumulative
distribution $P_N(S \le s)$ of the number of cycles in a directed 2-RRG 
of $N=10^4$ nodes,
obtained from Eq. (\ref{eq:PsdCum}). 
The analytical results are found to be in very good agreement with the
results obtained from computer simulations (circles).

\subsection{Distribution of the number of cycles in undirected 2-RRGs}

We now turn to calculate the distribution
$P_N(S=s)$
of the number of cycles
in undirected 2-RRGs.
Inserting the expression for $P_N( \{ g_{\ell} \} )$ from 
Eq. (\ref{eq:Pjd}) in Eq. (\ref{eq:PNSs0}), we obtain

\begin{equation}
P_N(S=s) = 
\frac{ (2N)!! }{ (2N-1)!! }
\sum_{g_1,\dots,g_N \ge 0} 
\prod_{\ell=1}^N
\frac{1}{(2 \ell)^{g_{\ell}} g_{\ell}!}  
\delta_{ \sum_{\ell} \ell g_{\ell},N }
\delta_{ \sum_{\ell} g_{\ell},s }.
\end{equation}

\noindent
Using the change of variables 
presented in Eq. (\ref{eq:covgell}),
we obtain

\begin{equation}
P_N(S=s) = 
\frac{ (2N)!! }{ (2N-1)!! }
\frac{1}{2^s s!}
\sum_{\ell_1,\dots,\ell_{s}=1}^N
\frac{1}{ \ell_1 \ell_2 \dots \ell_{s} }
\delta_{ \sum_i \ell_i,N }.
\label{eq:Pnc2}
\end{equation}

\noindent
Note that the sum on the right hand side of Eq. (\ref{eq:Pnc2})
is equal to $f_s(N)$, given by Eq. (\ref{eq:f(N)}).
We thus obtain

\begin{equation}
P_N(S=s) =   
\frac{ (2N)!! }{ (2N-1)!! }
\frac{1}{2^s N!}
\bigg[
\begin{array}{l}
N     \\
s    
\end{array} 
\bigg].
\label{eq:PNSsUD}
\end{equation}

\noindent
Below we show that the distribution $P_N(S=s)$ is properly normalized.
To this end we use equation 26.8.7 from Ref. \cite{Olver2010},
which can be written in the form

\begin{equation}
\sum_{s=1}^{N} (-1)^{N-s} 
\bigg[
\begin{array}{l}
N     \\
s    
\end{array} 
\bigg] x^s =
(x-N+1)_{N},
\label{eq:2687}
\end{equation}

\noindent
where $(a)_N$ is the Pochhammer symbol
\cite{Olver2010}.
Inserting $x=-1/2$ in Eq. (\ref{eq:2687}),
we obtain

\begin{equation}
\sum_{s=1}^{N} 
\bigg[
\begin{array}{l}
N     \\
s    
\end{array} 
\bigg]
\frac{1}{2^s} =
(-1)^{N} \left( \frac{1}{2} - N \right)_{N}.
\label{eq:PochN}
\end{equation}

\noindent
Expressing the Pochhammer symbol on the right hand side of Eq. (\ref{eq:PochN})
as a ratio between two Gamma functions, we obtain

\begin{equation}
\sum_{s=1}^{N} 
\bigg[
\begin{array}{l}
N     \\
s    
\end{array} 
\bigg]
\frac{1}{2^s} =
\frac{ (-1)^{N} \Gamma(1/2) }{\Gamma(1/2-N)}.
\label{eq:PochN2}
\end{equation}

\noindent
Using Euler's reflection formula
\cite{Olver2010}

\begin{equation}
\Gamma \left( \frac{1}{2} - N \right) \Gamma \left( \frac{1}{2} + N \right) =
(-1)^{N}  \pi,
\end{equation}

\noindent
and the Legendre duplication formula
\cite{Olver2010}

\begin{equation}
\Gamma(1/2+N) = 2^{1-2N} \sqrt{\pi} \frac{ \Gamma(2N) }{\Gamma(N)},
\end{equation}

\noindent
we obtain

\begin{equation}
\sum_{s=1}^{N} 
\bigg[
\begin{array}{l}
N     \\
s    
\end{array} 
\bigg]
\frac{1}{2^s} =
2^{1-2N} \frac{ \Gamma(2N) }{\Gamma(N)}.
\label{eq:PochN3}
\end{equation}

\noindent
Expressing the Gamma functions of integer variables in terms
of factorials and double factorials, we obtain

\begin{equation}
\sum_{s=1}^{N} 
\bigg[
\begin{array}{l}
N     \\
s    
\end{array} 
\bigg]
\frac{1}{2^s} =
\frac{ (2N-1)!! }{(2N)!!} N!. 
\label{eq:PochN4}
\end{equation}

\noindent
This confirms the normalization of $P_N(S=s)$, given by Eq. (\ref{eq:PNSsUD}).
From Eq. (\ref{eq:PNSsUD}) we obtain the cumulative distribution of the number of cycles,
which is given by

\begin{equation}
P_N(S \le s) =   
\frac{ (2N)!! }{ (2N-1)!! } 
\frac{1}{ N! }
\sum_{s'=1}^s
\frac{1}{2^{s'} }
\bigg[
\begin{array}{l}
N     \\
s'   
\end{array} 
\bigg].
\label{eq:PNSsUD_Cum}
\end{equation}

In Fig. \ref{fig:2}(b) we present exact
analytical results (solid line) for the cumulative
distribution $P(S \le s)$ of the number of cycles in undirected 2-RRGs 
that consist of $N=10$ nodes,
obtained from Eq. (\ref{eq:PNSsUD_Cum}). 
The analytical results are found to be in very good agreement with the
results obtained from computer simulations (circles).

While Eq. (\ref{eq:PNSsUD}) provides an exact result for $P_N(S=s)$, it is useful to 
express this distribution in terms of more elementary functions.
This is possible in the asymptotic limit of large $N$,
where the ratio of double factorials can be approximated by

\begin{equation}
\frac{(2N)!!}{(2N-1)!!} \simeq \sqrt {\pi N}+O\left( N^{-1/2}\right).
\end{equation}

\noindent
Using this result and inserting 
$f_s(N)$ from Eq. (\ref{eq:f(N)4}) into
Eq. (\ref{eq:Pnc2}), 
we find that

\begin{equation}
P_N(S=s) \simeq \frac{1}{2^s} 
\sqrt{\frac{\pi}{N}}
\sum_{i=1}^{s}
a_i \frac{ (\ln N)^{s - i} }{(s-i)!}.
\label{eq:Psu}
\end{equation}

\noindent
From Eq. (\ref{eq:Psu}) we obtain the cumulative distribution of the number of cycles,
which is given by

\begin{equation}
P_N(S \le s) \simeq 
\sqrt{\frac{\pi}{N}}
\sum_{s'=1}^s
\frac{1}{2^{s'}} 
\sum_{i=1}^{s'}
a_i \frac{ (\ln N)^{s' - i} }{(s' - i)!}.
\label{eq:PsuCum}
\end{equation}

In Fig. \ref{fig:3}(b) we present
analytical results (solid line) for the cumulative
distribution of the number of cycles in undirected 2-RRGs 
of $N=10^4$ nodes,
obtained from Eq. (\ref{eq:PsuCum}). 
The analytical results are found to be in very good agreement with the
results obtained from computer simulations (circles).

\section{Moments and cumulants}

In this section we calculate the moments and cumulants of 
the distribution of the number of cycles in 2-RRGs that consist of $N$ nodes.
To this end we introduce the moment generating function,
which is given by

\begin{equation}
M(t) = {\mathbb E} \left[ e^{t S} \right].
\label{eq:Mt}
\end{equation}

\noindent
The cumulant generating function is given by

\begin{equation}
K(t) = \ln M(t).
\label{eq:KtMt}
\end{equation}

\noindent
Using this function one can calculate the cumulants via
differentiation according to

\begin{equation}
\kappa_n = \frac{ d^n K(t) }{d t^n} \bigg\vert_{t=0}.
\label{eq:kappan}
\end{equation}

\subsection{Moments and cumulants in directed 2-RRGs}

The moment generating function of directed 2-RRGs 
is given by

\begin{equation}
M(t) = \frac{1}{N!} \sum_{s=0}^{N}
e^{t s} 
\bigg[
\begin{array}{l}
N     \\
s    
\end{array} 
\bigg].
\end{equation}

\noindent
Using Eq. (\ref{eq:2687}) with $x=-e^{t}$, we obtain

\begin{equation}
M(t) = \frac{ (-1)^N }{ N! }
\left( -e^t - N + 1 \right)_N.
\end{equation}

\noindent
The moment generating function $M(t)$ may also be written in
the form 

\begin{equation}
M(t) = \frac{ \Gamma(N+e^t) }{\Gamma(e^t) N!},
\end{equation}

\noindent
in agreement with the results presented in Ref. \cite{Shepp1966}.
The corresponding cumulant generating function is given by

\begin{equation}
K(t) =    \ln \left[ \frac{ \Gamma(N+e^t) }{\Gamma(e^t) N!} \right].
\label{eq:KD}
\end{equation}

\noindent
Using Eq. (\ref{eq:KD}) we obtain the first two cumulants, 
which are given by

\begin{equation}
\langle S \rangle = \kappa_1 = H_N,
\label{eq:k1D}
\end{equation}

\noindent
where $H_N$ is the harmonic number 
\cite{Boya2018},
and

\begin{equation}
{\rm Var}(S) = \kappa_2 = 
H_N - H_N^{(2)},
\label{eq:k2D}
\end{equation}

\noindent
where $H_N^{(m)}$ is the generalized harmonic number of order $m$
\cite{Boya2018}.
Similarly, one can calculate higher order cumulants such as

\begin{equation}
\kappa_3 = H_N - 3 H_N^{(2)} + 2 H_N^{(3)}
\label{eq:k3D}
\end{equation}

\noindent
and

\begin{equation}
\kappa_4 = H_N - 7 H_N^{(2)} + 12 H_N^{(3)} - 6 H_N^{(4)}.
\label{eq:k4D}
\end{equation}

\noindent
In the limit of large $N$ we can use the asymptotic expression for the
distribution $P_N(S=s)$,
given by Eq. (\ref{eq:Psd}),
and obtain

\begin{equation}
M(t) \simeq \frac{1}{N} \sum_{s=0}^{\infty}
\sum_{i=1}^{s} e^{s t} a_i \frac{ (\ln N)^{s-i} }{ (s-i)! }.
\end{equation}

\noindent
Exchanging the order of summations, we obtain

\begin{equation}
M(t) \simeq
\frac{1}{N} \sum_{i=1}^{\infty} a_i \sum_{s=i}^{\infty}
e^{s t} \frac{ (\ln N)^{s-i} }{ (s-i)! }.
\end{equation}

\noindent
Shifting the summation index in the second sum, we obtain

\begin{equation}
M(t) \simeq
\frac{1}{N} \sum_{i=1}^{\infty} a_i 
e^{i t}
\sum_{s=0}^{\infty}
e^{s t} \frac{ (\ln N)^{s} }{ s! }.
\end{equation}

\noindent
Using Eq. (\ref{eq:invGamma}) we carry out 
the two summations and obtain

\begin{equation}
M(t) \simeq \frac{1}{N} \frac{1}{\Gamma(e^t)} \exp \left( e^{t} \ln N \right).
\end{equation}

\noindent
Using Eq. (\ref{eq:KtMt}) we obtain the cumulant generating function,
which is given by

\begin{equation}
K(t) \simeq \left( e^t - 1 \right) \ln N - \ln \left[ \Gamma \left( e^t \right) \right].
\end{equation}

\noindent
Using Eq. (\ref{eq:kappan}) we obtain the cumulants, which take the form

\begin{equation}
\kappa_n \simeq \ln N - \frac{ d^n }{dt^n} \ln \left[ \Gamma \left( e^t \right) \right] \bigg\vert_{t=0}.
\end{equation}

\noindent
In order to calculate high order derivatives of
$\ln \left[ \Gamma \left( e^t \right) \right]$
we use the identity
(equation A.4 in Ref. \cite{Boya2018})

\begin{equation}
\frac{ d^n }{dt^n} f(e^t) = 
\sum_{m=1}^{n} 
\bigg\{
\begin{array}{l}
n     \\
m    
\end{array} 
\bigg\}
e^{mt} 
f^{(m)}(e^t),
\label{eq:dnfet}
\end{equation}

\noindent
where 
$\bigg\{
\begin{array}{l}
n     \\
m    
\end{array} 
\bigg\}$
is the Stirling number of the second kind
and $f^{(m)}(x)$ is the $m$th derivative of $f(x)$.
We also use the fact that

\begin{equation}
\frac{d^n}{dz^n} \ln \Gamma(z) = \psi^{(n-1)}(z),
\label{eq:dnzn}
\end{equation}

\noindent
where $\psi^{(n)}(z)$ is the $n$th derivative of the digamma function
\cite{Olver2010}.
Using these identities, we obtain

\begin{equation}
\frac{ d^n }{dt^n} \ln \left[ \Gamma \left( e^t \right) \right] \bigg\vert_{t=0} =
\sum_{m=1}^{n} 
\bigg\{
\begin{array}{l}
n     \\
m    
\end{array} 
\bigg\}
\psi^{(m-1)}(1).
\end{equation}

\noindent
It is also known that
(equation 5.4.12 in Ref. \cite{Olver2010})

\begin{equation}
\psi^{(0)}(1) = - \gamma,
\end{equation}

\noindent
and that for $m \ge 1$
(equation 5.15.2 in Ref. \cite{Olver2010})

\begin{equation}
\psi^{(m)}(1) =  (-1)^{m+1} m! \zeta(m+1),
\end{equation}

\noindent
where $\zeta(m)$ is the Riemann zeta function
\cite{Olver2010}.
Combining the results derived above, we obtain

\begin{equation}
\kappa_n \simeq \ln N + \gamma + 
\sum_{m=2}^{n}
\bigg\{
\begin{array}{l}
n     \\
m    
\end{array} 
\bigg\}
(- 1)^{m-1} (m-1)! \zeta(m).
\label{eq:kappa_nd}
\end{equation}

\noindent
Using Eq. (\ref{eq:kappa_nd}) we write down explicitly the first few cumulants of
$P_N(S=s)$
in the large $N$ limit. 
They are given by

\begin{eqnarray}
\kappa_1 &\simeq& \ln N + \gamma
\nonumber \\
\kappa_2 &\simeq& \ln N + \gamma - \frac{ \pi^ 2}{6}
\nonumber \\
\kappa_3 &\simeq& \ln N + \gamma - \frac{ \pi^2 }{ 2 } + 2 \zeta(3)
\nonumber \\
\kappa_4 &\simeq& \ln N + \gamma - \frac{ 7 \pi^2 }{ 6 } + 12 \zeta(3) - \frac{\pi^4}{15}.
\end{eqnarray}

\noindent
The results for $\kappa_1$ and $\kappa_2$ are in agreement
with the classical results reported in Refs.
\cite{Goncharov1942,Goncharov1944,Greenwood1953,Wilf1990}.
In the large $N$ limit all the cumulants are of the form $\ln N + {\cal O}(1)$.
This essentially implies that in the large $N$ limit the distribution 
$P_N(S=s)$ approaches a Poisson distribution with a parameter $\ln N$.

Comparing between Eqs. (\ref{eq:k1D})-(\ref{eq:k4D}) 
and Eq. (\ref{eq:kappa_nd}), and using
the fact that for $m \ge 2$

\begin{equation}
\lim_{N \rightarrow \infty} H_N^{(m)} = \zeta(m),
\end{equation}

\noindent
we obtain a general expression for the cumulants at
finite values of $N$, which is given by

\begin{equation}
\kappa_n = H_N + 
\sum_{m=2}^{n}
\bigg\{
\begin{array}{l}
n     \\
m    
\end{array} 
\bigg\}
(- 1)^{m-1} (m-1)! H_N^{(m)}.
\label{eq:kappa_nd2}
\end{equation}

\begin{figure}
\begin{center}
\includegraphics[width=7cm]{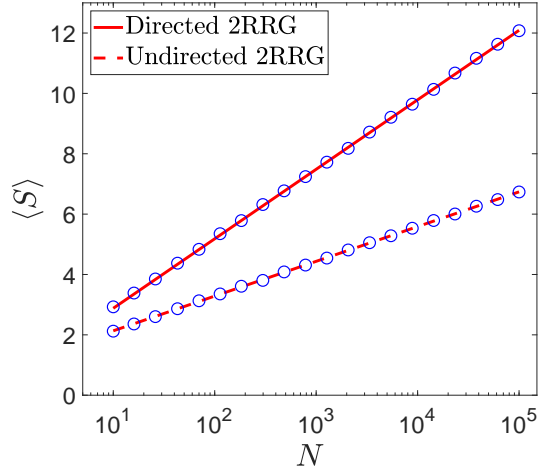} 
\end{center}
\caption{
Analytical results for the 
mean number of cycles $\langle S \rangle$ as a function of the network size $N$,
in directed 2-RRGs (solid line)
and in undirected 2-RRGs (dashed line),
obtained from Eqs. (\ref{eq:k1D}) and (\ref{eq:k1U}), 
respectively. The analytical results are in very good agreement with
the results obtained from computer simulations (circles).
To leading order, in directed 2-RRGs $\langle S \rangle \simeq \ln N$,
while in undirected 2-RRGs $\langle S \rangle \simeq \frac{1}{2} \ln N$.
}
\label{fig:4}
\end{figure}

In Fig. \ref{fig:4} we present
analytical results (solid line) for the 
mean number of cycles $\langle S \rangle$ in directed 2-RRGs
as a function of the network size $N$,
obtained from Eq. (\ref{eq:k1D}).
The analytical results are in very good agreement with
the results obtained from computer simulations (circles).

\begin{figure}
\begin{center}
\includegraphics[width=7cm]{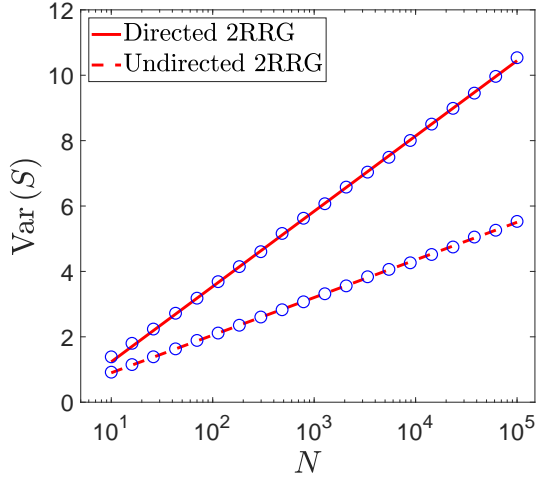} 
\end{center}
\caption{
Analytical results for the 
variance of the distribution of the number of cycles 
as a function of the network size $N$,
in directed 2-RRGs (solid line)
and in undirected 2-RRGs (dashed line),
obtained from Eqs. (\ref{eq:k2D}) and (\ref{eq:k2U}), 
respectively. The analytical results are in very good agreement with
the results obtained from computer simulations (circles).
To leading order, in directed 2-RRGs ${\rm Var}(S) \simeq \ln N$,
while in undirected 2-RRGs ${\rm Var}(S) \simeq \frac{1}{2} \ln N$.
}
\label{fig:5}
\end{figure}

In Fig. \ref{fig:5} we present
analytical results (solid line) for the 
variance ${\rm Var}(S)$
in directed 2-RRGs 
as a function of the network size $N$,
obtained from Eq. (\ref{eq:k2D}).
The analytical results are in very good agreement with
the results obtained from computer simulations (circles).

\subsection{Moments and cumulants in undirected 2-RRGs}
 
 The moment generating function of undirected 2-RRGs 
is given by

\begin{equation}
M(t) = 
\frac{ (2N)!! }{ (2N - 1)!! }
\frac{1}{N!} \sum_{s=0}^{N}
\frac{ e^{t s} }{ 2^s }
\bigg[
\begin{array}{l}
N     \\
s    
\end{array} 
\bigg].
\end{equation}

\noindent
Using Eq. (\ref{eq:2687}) with $x=-e^{t}/2$, we obtain

\begin{equation}
M(t) = \frac{ (-1)^N }{ N! }
\frac{ (2N)!! }{ (2N-1)!! }
\left( - \frac{e^t}{2} - N + 1 \right)_N.
\end{equation}

\noindent
The moment generating function $M(t)$ may also be written in
the form

\begin{equation}
M(t) = \frac{ (2N)!! }{ (2N-1)!! }
\frac{ \Gamma \left( N+ \frac{e^t}{ 2} \right) }{\Gamma \left( \frac{e^t}{2} \right) N!}.
\end{equation}

\noindent
The corresponding cumulant generating function is given by

\begin{equation}
K(t) =    \ln \left[
\frac{ (2N)!! }{ (2N-1)!! }
 \frac{ \Gamma \left( N+ \frac{e^t}{ 2} \right) }
{\Gamma \left( \frac{e^t}{2} \right) N!} \right].
\end{equation}

\noindent
Using Eq. (\ref{eq:kappan}) we obtain the first four cumulants.
The first cumulant is given by

\begin{equation}
\langle S \rangle = \kappa_1 = 
\frac{1}{2} H_{N-\frac{1}{2}} + \ln 2,
\label{eq:k1U}
\end{equation}

\noindent
where $H_{N-\frac{1}{2}}$ is an Harmonic number at a half-integer value
\cite{Sofo2016}.
The second cumulant is given by

\begin{equation}
{\rm Var}(S) = \kappa_2 = 
\frac{1}{2} H_{N-\frac{1}{2}} + \ln 2 
- \frac{1}{4} \left[ H_{N-\frac{1}{2}}^{(2)}
+ 2 \zeta(2) \right],
\label{eq:k2U}
\end{equation}

\noindent
while the third and fourth cumulants are given by

\begin{equation}
\kappa_3 = 
\frac{1}{2} H_{N-\frac{1}{2}} + \ln 2 
- \frac{3}{4} \left[ H_{N-\frac{1}{2}}^{(2)}
+ 2 \zeta(2) \right]
+ \frac{1}{4} \left[ H_{N-\frac{1}{2}}^{(3)} + 6 \zeta(3) \right]
\label{eq:k3U}
\end{equation}

\noindent
and

\begin{equation}
\kappa_4 = 
\frac{1}{2} H_{N-\frac{1}{2}} + \ln 2 
- \frac{7}{4} \left[ H_{N-\frac{1}{2}}^{(2)}
+ 2 \zeta(2) \right]
+ \frac{3}{2} \left[ H_{N-\frac{1}{2}}^{(3)} + 6 \zeta(3) \right]
- \frac{3}{8} \left[ H_{N-\frac{1}{2}}^{(4)} + 14 \zeta(4) \right].
\label{eq:k4U}
\end{equation}

In the limit of large $N$ we can use the asymptotic expression for the
distribution $P_N(S=s)$, 
given by Eq. (\ref{eq:Psu}).
Inserting it into Eq. (\ref{eq:Mt}) we obtain
an asymptotic expression for the moment generating function,
which is given by

\begin{equation}
M(t) \simeq \sqrt{ \frac{\pi}{N} } \sum_{s=0}^{\infty}
\sum_{i=1}^{s} e^{s t} \frac{ a_i }{ 2^i } \frac{ (\ln N)^{s-i} }{ 2^{s-i} (s-i)! }.
\end{equation}

\noindent
Exchanging the order of summations and shifting the summation
index in the second sum, we obtain

\begin{equation}
M(t) \simeq
\sqrt{ \frac{\pi}{N} } \sum_{i=1}^{\infty} a_i 
\frac{ e^{i t} }{2^i}
\sum_{s=0}^{\infty}
e^{s t} \frac{ (\ln N)^{s} }{2^s s! }.
\end{equation}

\noindent
Using Eq. (\ref{eq:invGamma}) we carry out 
the two summations and obtain

\begin{equation}
M(t) \simeq \sqrt{ \frac{\pi}{N} } \frac{1}{\Gamma(e^t/2)} 
\exp \left( \frac{ e^{t} }{2} \ln N \right).
\end{equation}

\noindent
Using Eq. (\ref{eq:KtMt}) we obtain the cumulant generating function,
which is given by

\begin{equation}
K(t) \simeq   \frac{ e^t - 1 }{2}   \ln N - \ln \left[ \frac{1}{\sqrt{\pi}} 
\Gamma \left( \frac{ e^t }{2} \right) \right].
\end{equation}

\noindent
Using Eq. (\ref{eq:kappan}) we obtain the cumulants, which take the form

\begin{equation}
\kappa_n \simeq \frac{1}{2} \ln N 
- \frac{ d^n }{dt^n} \ln \left[ \frac{1}{\sqrt{\pi}}
\Gamma \left( \frac{e^t}{2} \right) \right] \bigg\vert_{t=0}.
\end{equation}

\noindent
Using Eqs. (\ref{eq:dnfet}) and (\ref{eq:dnzn}), we obtain

\begin{equation}
\frac{ d^n }{dt^n} \ln \left[ \frac{1}{\sqrt{\pi}} 
\Gamma \left( \frac{ e^t }{ 2 } \right) \right] \bigg\vert_{t=0} =
\sum_{m=1}^{n} 
\bigg\{
\begin{array}{l}
n     \\
m    
\end{array} 
\bigg\}
2^{-m}
\psi^{(m-1)} \left( \frac{1}{ 2} \right).
\end{equation}

\noindent
It is also known that
\cite{Choi2007}

\begin{equation}
\psi^{(0)} \left( \frac{1}{2} \right) = - \gamma - \ln 4,
\end{equation}

\noindent
and that for $m \ge 1$
\cite{Choi2007}

\begin{equation}
\psi^{(m)} \left( \frac{1}{2} \right) =  (-1)^{m+1} m!  \left( 2^{m+1} - 1 \right) \zeta(m+1).
\end{equation}

\noindent
Combining the results derived above, we obtain

\begin{equation}
\kappa_n \simeq \frac{ \ln N }{2} + \frac{ \gamma + \ln 4 }{2} +
\sum_{m=2}^{n}
\bigg\{
\begin{array}{l}
n     \\
m    
\end{array} 
\bigg\}
(- 1)^{m-1}  \left( 1 - 2^{-m} \right) (m-1)! \zeta(m),
\label{eq:kappa_nu}
\end{equation}

\noindent
which becomes exact in the large $N$ limit.
Using Eq. (\ref{eq:kappa_nu}) we write down explicitly the first few cumulants of
$P_N(S=s)$. They are given by

\begin{eqnarray}
\kappa_1 &\simeq& \frac{ \ln N }{2} + \frac{ \gamma + \ln 4 }{2}
\nonumber \\
\kappa_2 &\simeq& \frac{ \ln N }{2} + \frac{ \gamma + \ln 4 }{2} - \frac{ \pi^ 2}{8}
\nonumber \\
\kappa_3 &\simeq& \frac{ \ln N }{2} + \frac{ \gamma + \ln 4 }{2} 
- \frac{3 \pi^2 }{ 8 } + \frac{7}{4} \zeta(3)
\nonumber \\
\kappa_4 &\simeq& \frac{ \ln N }{2} + \frac{ \gamma + \ln 4 }{2} 
- \frac{7 \pi^2 }{ 8 } + \frac{21}{2} \zeta(3) - \frac{\pi^4}{16}.
\end{eqnarray}

\noindent
In the large $N$ limit all the cumulants are of the form $\frac{1}{2} \ln N + {\cal O}(1)$.
This essentially implies that in the large $N$ limit the distribution 
$P_N(S=s)$ approaches a Poisson distribution with a parameter $\frac{1}{2} \ln N$.

Using the fact that for $m \ge 2$

\begin{equation}
\lim_{N \rightarrow \infty} H_{N-\frac{1}{2}}^{(m)} = \zeta(m),
\end{equation}

\noindent
we obtain a general expression for the cumulants at
finite values of $N$, which is given by

\begin{equation}
\kappa_n = \frac{1}{2} H_{N-\frac{1}{2}} + \ln 2 + 
\sum_{m=2}^{n}
\bigg\{
\begin{array}{l}
n     \\
m    
\end{array} 
\bigg\}
(- 1)^{m-1} 2^{-m} (m-1)! \left[ H_{N-\frac{1}{2}}^{(m)}
+ (2^m-2) \zeta(m) \right].
\label{eq:kappa_nud2}
\end{equation}

In Fig. \ref{fig:4} we present
analytical results (dashed line) for the 
mean number of cycles $\langle S \rangle$ in undirected 2-RRGs 
as a function of the network size $N$,
obtained from Eq. (\ref{eq:k1U}).
The analytical results are in very good agreement with
the results obtained from computer simulations (circles).

In Fig. \ref{fig:5} we present
analytical results (dashed line) for the 
variance ${\rm Var}(S)$ in undirected 2-RRGs 
as a function of the network size $N$,
obtained from Eq. (\ref{eq:k2U}).
The analytical results are in very good agreement with
the results obtained from computer simulations (circles).

 \section{Discussion}
 
Below we discuss the similarities and differences between the
directed and undirected 2-RRGs.
In directed 2-RRGs the
mean number of
cycles $\langle S \rangle$   
scales with $\ln N$, while in undirected 2-RRGs it scales with
$\frac{1}{2} \ln N$.
Thus, the expected number of cycles in directed 2-RRGs is
twice as large as in undirected 2-RRGs.
This is due to the fact that in the construction of undirected 2-RRGs 
each end of a given chain may connect to both sides of any other
linear chain, while in directed 2-RRGs it may only connect to the
complementary side.
As a result, in undirected 2-RRGs the connection of chains
forming a longer chain is more probable than in directed 2-RRGs.
Thus, in undirected 2-RRGs the competing process of closing a chain to form a
cycle is less probable than in directed 2-RRGs.
This implies that in undirected 2-RRGs the cycles are expected
to be longer and their number is expected to be
smaller than in directed 2-RRGs.
 
2-RRGs are marginal networks that reside at the boundary between 
the subcritical regime and the supercritical regime.
In the subcritical regime, configuration model networks consist of 
many finite tree components. The distribution of sizes of these tree
components can be calculated using the framework of generating
functions
\cite{Newman2007}.
In this framework it is assumed that all the network components
exhibit a tree structure. In the 2-RRG the topological constraint 
that all the nodes are of degree $k=2$ imposes the formation
of cycles. Therefore, the generating function formalism cannot be used
to analyze the distribution of cycle lengths in 2-RRGs.
A naive attempt to use the generating function formalism to obtain
the distribution of cluster sizes (which are also the cycle lengths)
in 2-RRGs fails to determine the distribution.

RRGs with $c \ge 3$ are supercritical. 
They consist of a giant component that encompasses the whole network.
While the local structure of the the network is typically tree-like, at larger scales
it exhibits cycles with a broad distribution of cycle lengths.
The length of a cycle is given by the number of nodes (or edges) that reside
along the cycle. 
The longest possible cycle is a Hamiltonian cycle of length $\ell=N$.
The expected number of cycles
of length $\ell$ in an undirected RRG that consists of $N$ nodes of degree $c \ge 3$,
where $\ell \ll \ln N$,
is given by 
\cite{Marinari2004,Bianconi2005,Marinari2006}
 
\begin{equation}
\langle G_{\ell} \rangle =  \frac{ (c-1)^{\ell} }{  2 \ell  }.
\label{eq:Gell1}
\end{equation}

\noindent
This implies that for $c \ge 3$ the number of cycles of length $\ell$ proliferates
exponentially as $\ell$ is increased, as long as $\ell \ll \ln N$.
Although these results were not claimed to hold in the case of $c=2$,
it is interesting to examine their relevance to 2-RRGs.
In the special case of an undirected 2-RRG, where $c=2$, Eq. (\ref{eq:Gell1})  is reduced to

\begin{equation}
\langle G_{\ell} \rangle = \frac{ 1 }{2 \ell}.
\label{eq:Gell2}
\end{equation}

\begin{figure}
\begin{center}
\includegraphics[width=7cm]{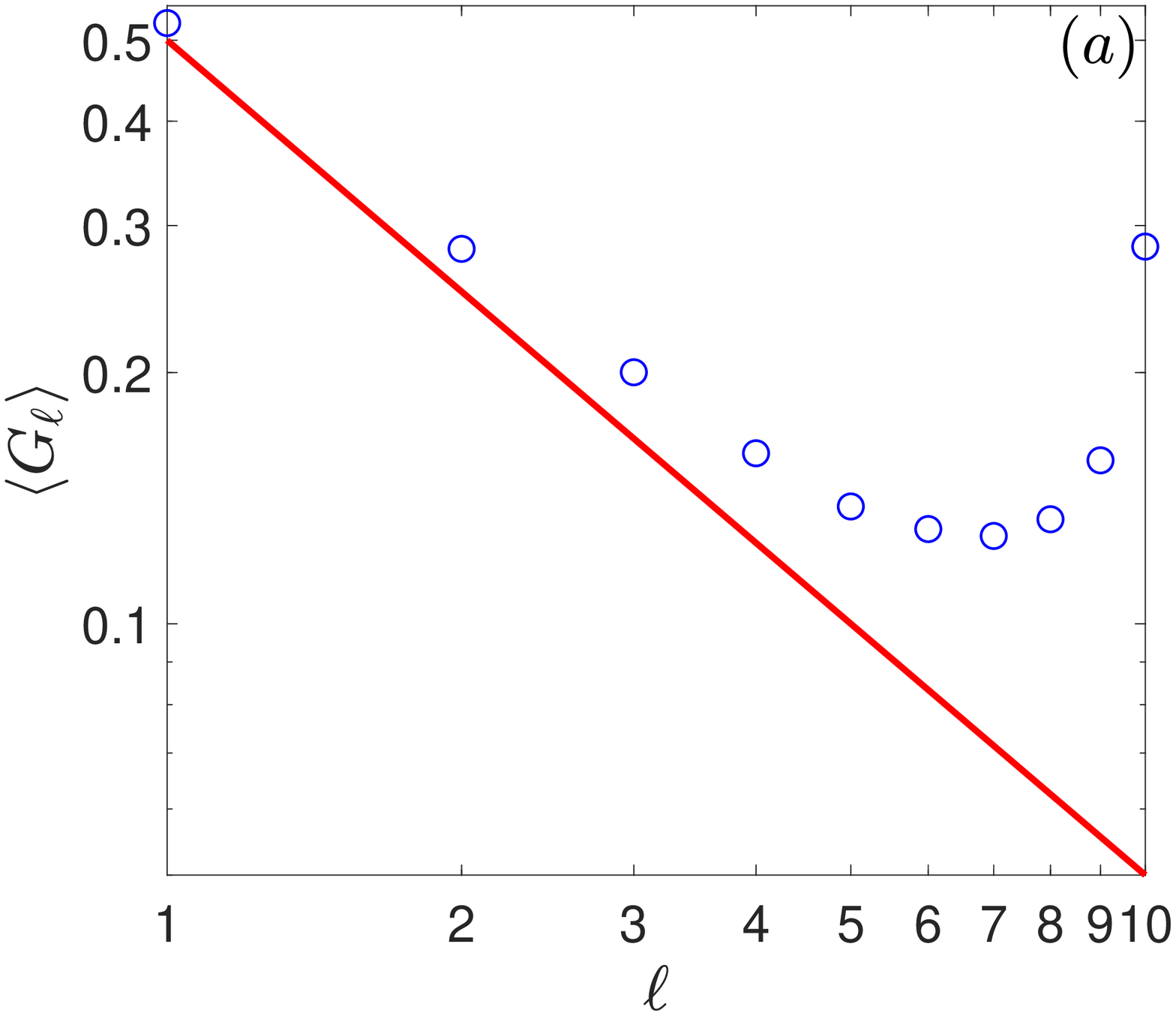} 
\hspace{0.5in}
\includegraphics[width=7cm]{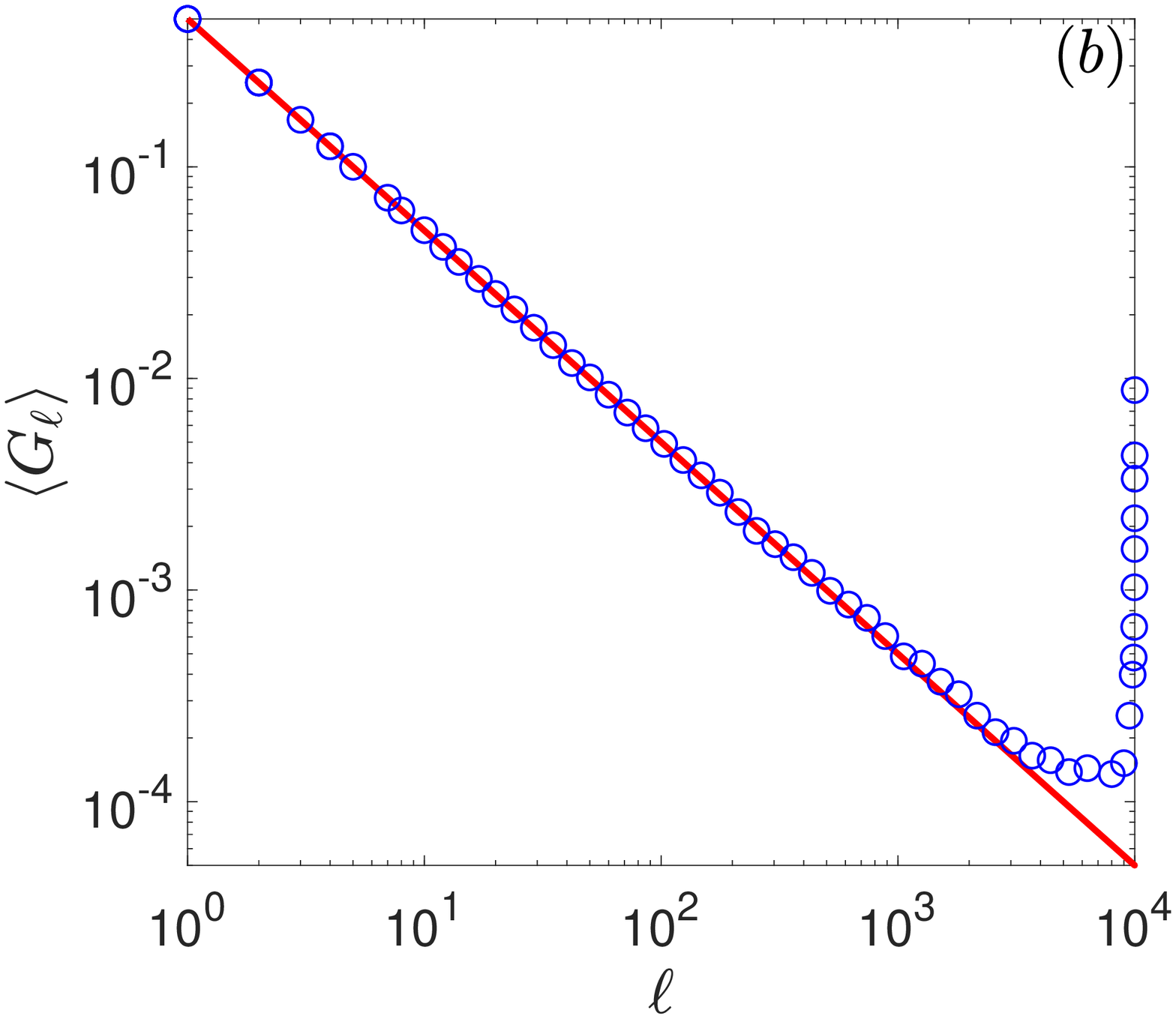}
\end{center}
\caption{
Analytical results (solid lines) for the expected number $\langle G_{\ell} \rangle$
of cycles of length $\ell$,
in an undirected 2-RRG that consists of $N$ nodes, for $N=10$ (a) and for
$N=10^4$ (b), obtained from Eq. (\ref{eq:Gell2}), on a log-log scale.
We also present results obtained from computer simulations (circles).
For $N=10$ the simulation results deviate significantly from the
prediction of Eq. (\ref{eq:Gell2}). For $N=10^4$ there is a very good
agreement between the analytical results and the simulation results
in the range of $\ell \ll N$. 
The agreement between the analytical results and the simulation
results improves as $N$ is increased.
}
\label{fig:6}
\end{figure}

\noindent
In Fig. \ref{fig:6} we present analytical results (solid lines)
for the expected number $\langle G_{\ell} \rangle$
of cycles of length $\ell$ in undirected 2-RRGs, obtained from Eq. (\ref{eq:Gell2}),
as a function of $\ell$ for $N=10$ (a) and $N=10^4$ (b).
We also present the results obtained from computer simulations (circles).
It is found that for $N=10$ there is a big difference between the
analytical results obtained from Eq. (\ref{eq:Gell2}) and the simulation results.
In contrast, for $N=10^4$ the analytical results are in very good agreement
with the results of computer simulations for $\ell \ll N$.
This implies that Eq. (\ref{eq:Gell2}) is valid for 2-RRGs in the
large network limit and for sufficiently short cycles.
For larger values of $\ell$ Eq. (\ref{eq:Gell2}) is no longer valid,
as $\langle G_{\ell} \rangle$ becomes an increasing function of $\ell$.
Note that the simulation results for $\langle G_{\ell} \rangle$
exceed the values predicted by 
Eq. (\ref{eq:Gell2}).
The total number of nodes can be expressed 
in the form

\begin{equation}
N = \sum_{\ell =1}^{N} \ell \langle G_{\ell} \rangle,
\label{eq:NGell}
\end{equation}

\noindent
which is obtained by averaging Eq. (\ref{eq:sumN}) over the ensemble.
Inserting $\langle G_{\ell} \rangle$
from Eq. (\ref{eq:Gell2}) into the right hand side of Eq. (\ref{eq:NGell}),
it yields only $N/2$ nodes instead of $N$ nodes.
This implies that Eq. (\ref{eq:Gell2}) is valid only as long as $\ell \ll N$.
Indeed, Fig. \ref{fig:6} reveals that Eq. (\ref{eq:Gell2})
misses the very long cycles whose length is of order $N$.

\section{Summary}

2-RRGs are networks in which each node has two links.
Therefore, these networks consist of
a set of closed cycles whose lengths are determined by the
random process of bond formation between the nodes.
In this paper we have calculated the distributions $P_N(S=s)$
of the number of cycles
in directed and undirected 2-RRGs.
Starting from the
joint distributions of cycle lengths 
$P_N(\{ g_{\ell} \})$
we obtained exact results for $P_N(S=s)$, which are
expressed in terms of the Stirling numbers of the first kind.
In sufficiently large networks these distributions can be expressed in terms of
more elementary functions.
We also derived closed-form expressions 
for the moments and cumulants of 
$P_N(S=s)$.
It was found that to
leading order, in directed 2-RRGs,
the cumulants of all orders $n=1,2,\dots$ satisfy
$\kappa_n \simeq \ln N$,
while in undirected 2-RRGs they satisfy
$\kappa_n \simeq  \frac{1}{2} \ln N$.
This implies that in the large $N$ limit the distributions $P_N(S=s)$
converge towards the Poisson distribution. 

This work was supported by the Israel Science Foundation grant no. 1682/18.




\begin{thebibliography}{10}

\bibitem{Bollobas2001}
B. Bollob\'as, 
{\it Random Graphs, Second Edition} 
(Cambridge University Press, Cambridge, 2001).


\bibitem{Dorogovtsev2003}
S.N. Dorogovtsev and J.F.F. Mendes,
{\it Evolution of Networks: From Biological Nets to the Internet and WWW}
(Oxford University Press, Oxford, 2003).



\bibitem{Havlin2010}
S. Havlin and R. Cohen,
{\it Complex Networks: Structure, Robustness and Function}
(Cambridge University Press, New York, 2010).


\bibitem{Estrada2011}
E. Estrada, 
{\it The structure of complex networks: Theory and applications} 
(Oxford University Press, Oxford, 2011).


\bibitem{Barrat2012}
A. Barrat, M. Barth\'elemy and A. Vespignani,
{\it Dynamical Processes on Complex Networks}
(Cambridge University Press, Cambridge, 2012).




\bibitem{Latora2017}
V. Latora, V. Nicosia and G. Russo,
{\it Complex Networks: Principles, Methods and Applications}
(Cambridge University Press, Cambridge, 2017).
 


\bibitem{Newman2018}
M.E.J. Newman, 
{\it Networks: an Introduction, Second Edition} 
(Oxford University Press, Oxford, 2018).

\bibitem{Hofstad2016}
R. van der Hofstad,
{\it Random graphs and complex networks}
(Cambridge University Press, Cambridge, 2016).

\bibitem{Dorogovtsev2022}
S.N. Dorogovtsev and J.F.F. Mendes,
{\it The Nature of Complex Networks}
(Oxford University Press, Oxford, 2022).




\bibitem{Bollobas1980}
B. Bollob\'as,
A probabilistic proof of an asymptotic formula for the number of 
labelled regular graphs,
{\it Euro. J. Combin.} {\bf 1}, 311 (1980).


\bibitem{Molloy1995}
M. Molloy and A. Reed,
A critical point for random graphs with a given degree sequence,
{\it Random Structures and Algorithms} {\bf 6}, 161 (1995).

\bibitem{Molloy1998}
M. Molloy and A. Reed,
The size of the giant component of a random graph 
with a given degree sequence,
{\it Combin., Prob. and Comp.} {\bf 7}, 295 (1998).


\bibitem{Newman2001}
M.E.J. Newman, S.H. Strogatz and D.J. Watts,
Random graphs with arbitrary degree distributions
and their applications,
{\it Phys. Rev. E} {\bf 64}, 026118 (2001).





\bibitem{Fosdick2018}
B.K. Fosdick, D.B. Larremore, J. Nishimura and J. Ugander, 
Configuring random graph models with fixed degree sequences,
{\it SIAM Review} {\bf 60}, 315 (2018).



\bibitem{Arratia1992}
R. Arratia and S. Tavar\'e,
The cycle structure of random permutations,
{\it The Annals of Probability} {\bf 20}, 1567 (1992).

\bibitem{Shepp1966}
L.A. Shepp and S.P. Lloyd,
Ordered cycle lengths in a random permutation,
{\it Transactions of the American Mathematical Society} {\bf 121}, 340 (1966).

\bibitem{Flajolet1990}
P. Flajolet and A.M. Odlyzko,
Random mapping statistics,
Advances in Cryptography,
J.J. Quisquater and J. Vandewalle (Eds.)
Eurocrypt '89, LNCS 434, 329 (1990)


\bibitem{Golomb1964}
S. W. Golomb, 
Random permutations,
{\it Bull. Amer. Math. Soc.} {\bf 70}, 747 (1964).

\bibitem{Golomb2017}
S. W. Golomb, 
{\it Shift Register Sequences, Third Edition}
(World Scientific, Singapore, 2017).	


\bibitem{Bona2012}
M. B\'ona,
{\it Combinatorics of Permutations, Second Edition}
(CRC Press, Boca Raton, 2012).



\bibitem{Dickman1930}
K. Dickman, 
On the frequency of numbers containing prime 
factors of a certain relative magnitude,
{\it Ark. Mat. Astron. Fysik} {\bf 22A}, 1 (1930).




\bibitem{Finch2003}
S. R. Finch,
Mathematical Constants,
(Cambridge University Press, Cambridge, 2003).


\bibitem{Golomb1998}
S. W. Golomb and P. Gaal, 
On the number of permutations on $n$ objects with greatest cycle length $k$,
{\it Adv. Appl. Math.} {\bf 20}, 98 (1998).


\bibitem{Knuth1976}
D. E. Knuth and L. Trabb Pardo, 
Analysis of a simple factorization algorithm, 
{\it Theoret. Comput. Sci.} {\bf 3}, 321 (1976).






\bibitem{Olver2010}
F.W.J. Olver, D.M. Lozier, R.F. Boisvert and C.W. Clark, 
{\it NIST handbook of mathematical functions} 
(Cambridge University Press, Cambridge, 2010).


\bibitem{Phillips2015}
C.L. Phillips, H.T. Nagle and A. Chakrabortty,
{\it Digital Control System: Analysis and Design, Fourth Edition}
(Pearson Education, Harlow, 2015).



\bibitem{Schlomilch1852}
O. Schl\"omilch, 
Recherches sur les coefficients des facult\'es analytiques,
{\it Crelle} {\bf 44}, 344 (1852).

\bibitem{Titchmarsh1939}
E.C. Titchmarsh, 
{\it The Theory of Functions, Second Edition}
(Oxford University Press, Oxford, 1939).

\bibitem{Wrench1968}
J.W. Wrench,
Concerning two series for the gamma function,
{\it Mathematics of Computation} {\bf 22}, 617 (1968).

\bibitem{Wrench1973}
J.W. Wrench, Erratum: Concerning two series for the gamma function,
{\it Mathematics of Computation} {\bf 27}, 681 (1973).

\bibitem{Ahmed2014}
L. Fekih-Ahmed, On the Power Series Expansion of the Reciprocal Gamma Function,
HAL archives, https://hal.archives-ouvertes.fr/hal-01029331v1, arXiv:1407.5983.




\bibitem{Boya2018}
K.N. Boyadzhiev,
{\it Notes on the Binomial Transform}
(World Scientific, Singapore, 2018).


\bibitem{Goncharov1942}
W. Goncharov, Sur la distribution des cycles dans les permutations, 
{\it C. R. (Dokl.) Acad. Sci. URSS} {\bf 35}, 267 (1942).

\bibitem{Goncharov1944}
W. Goncharov, On the field of combinatory analysis, 
{\it Soviet Math. Izv., Ser. Math} {\bf 8}, 3 (1944).

\bibitem{Greenwood1953}
R. E. Greenwood, The number of cycles associated with the elements of a permutation group, 
{\it Amer. Math. Monthly} {\bf 60}, 407 (1953).

\bibitem{Wilf1990}
H. S. Wilf, 
{\it generatingfunctionology, Third Edition} (A K Peters, Wellesley, 2005).



\bibitem{Sofo2016}
A. Sofo, 
Hamonic numbers at half-integer values,
{\it Integral Transforms and Special Functions} {\bf 27}, 430 (2016).


\bibitem{Choi2007}
J. Choi and D. Cvijovi\'c,
Values of the polygamma functions at rational arguments,
{\it J. Phys. A} {\bf 40}, 15019 (2007).

 
\bibitem{Newman2007}
M.E.J. Newman,
Component sizes in networks with arbitrary degree distributions,
{\it Phys. Rev. E} {\bf 76}, 045101 (2007).
 

\bibitem{Marinari2004}
E. Marinari and R. Monasson, 
Circuits in random graphs: from local trees to global loops,
{\it J. Stat. Mech.}, P09004 (2004). 

\bibitem{Bianconi2005}
G. Bianconi and M. Marsili,
Loops of any size and Hamilton cycles in random scale-free networks
{\it J. Stat. Mech.}, P06005 (2005).


\bibitem{Marinari2006}
E. Marinari and G. Semerjian, 
On the number of circuits in random graphs,
{\it J. Stat. Mech.}, P06019 (2006).



\end{thebibliography}
\end{document}